\documentclass[twoside,twocolumn,9pt]{article}
\usepackage{extsizes}
\usepackage[super,sort&compress,comma]{natbib} 
\usepackage[version=3]{mhchem}
\usepackage[left=1.5cm, right=1.5cm, top=1.785cm, bottom=2.0cm]{geometry}
\usepackage{balance}
\usepackage{mathptmx}
\usepackage{sectsty}
\usepackage{graphicx} 
\usepackage{lastpage}
\usepackage{color,soul}
\usepackage[format=plain,justification=justified,singlelinecheck=false,font={stretch=1.125,small,sf},labelfont=bf,labelsep=space]{caption}
\usepackage{float}
\usepackage{fancyhdr}
\usepackage{fnpos}
\usepackage[english]{babel}
\addto{\captionsenglish}{%
  
}
\usepackage{array}
\usepackage{droidsans}
\usepackage{charter}
\usepackage[T1]{fontenc}
\usepackage[usenames,dvipsnames]{xcolor}
\usepackage{setspace}
\usepackage[compact]{titlesec}
\usepackage{hyperref}

\usepackage{epstopdf}

\definecolor{cream}{RGB}{222,217,201}

\begin{document}

\pagestyle{fancy}
\thispagestyle{plain}
\fancypagestyle{plain}{
\renewcommand{\headrulewidth}{0pt}
}

\makeFNbottom
\makeatletter
\renewcommand\LARGE{\@setfontsize\LARGE{15pt}{17}}
\renewcommand\Large{\@setfontsize\Large{12pt}{14}}
\renewcommand\large{\@setfontsize\large{10pt}{12}}
\renewcommand\footnotesize{\@setfontsize\footnotesize{7pt}{10}}
\makeatother

\renewcommand{\thefootnote}{\fnsymbol{footnote}}
\renewcommand\footnoterule{\vspace*{1pt}%
\color{cream}\hrule width 3.5in height 0.4pt \color{black}\vspace*{5pt}} 
\setcounter{secnumdepth}{5}

\makeatletter 
\renewcommand\@biblabel[1]{#1}            
\renewcommand\@makefntext[1]%
{\noindent\makebox[0pt][r]{\@thefnmark\,}#1}
\makeatother 
\renewcommand{\figurename}{\small{Fig.}~}
\sectionfont{\sffamily\Large}
\subsectionfont{\normalsize}
\subsubsectionfont{\bf}
\setstretch{1.125} 
\setlength{\skip\footins}{0.8cm}
\setlength{\footnotesep}{0.25cm}
\setlength{\jot}{10pt}
\titlespacing*{\section}{0pt}{4pt}{4pt}
\titlespacing*{\subsection}{0pt}{15pt}{1pt}

\fancyfoot{}
\fancyfoot[LO,RE]{\vspace{-7.1pt}\includegraphics[height=9pt]{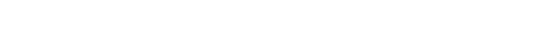}}
\fancyfoot[CO]{\vspace{-7.1pt}\hspace{13.2cm}\includegraphics{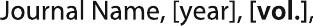}}
\fancyfoot[CE]{\vspace{-7.2pt}\hspace{-14.2cm}\includegraphics{head_foot/RF}}
\fancyfoot[RO]{\footnotesize{\sffamily{1--\pageref{LastPage} ~\textbar  \hspace{2pt}\thepage}}}
\fancyfoot[LE]{\footnotesize{\sffamily{\thepage~\textbar\hspace{3.45cm} 1--\pageref{LastPage}}}}
\fancyhead{}
\renewcommand{\headrulewidth}{0pt} 
\renewcommand{\footrulewidth}{0pt}
\setlength{\arrayrulewidth}{1pt}
\setlength{\columnsep}{6.5mm}
\setlength\bibsep{1pt}

\makeatletter 
\newlength{\figrulesep} 
\setlength{\figrulesep}{0.5\textfloatsep} 

\newcommand{\topfigrule}{\vspace*{-1pt}%
\noindent{\color{cream}\rule[-\figrulesep]{\columnwidth}{1.5pt}} }

\newcommand{\botfigrule}{\vspace*{-2pt}%
\noindent{\color{cream}\rule[\figrulesep]{\columnwidth}{1.5pt}} }

\newcommand{\dblfigrule}{\vspace*{-1pt}%
\noindent{\color{cream}\rule[-\figrulesep]{\textwidth}{1.5pt}} }

\makeatother

\twocolumn[
  \begin{@twocolumnfalse}
{\includegraphics[height=30pt]{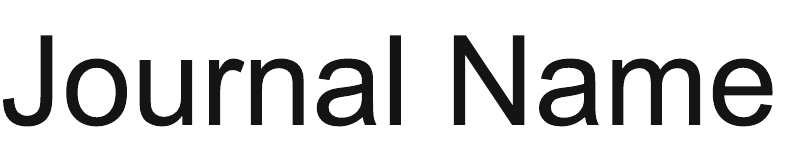}\hfill\raisebox{0pt}[0pt][0pt]{\includegraphics[height=55pt]{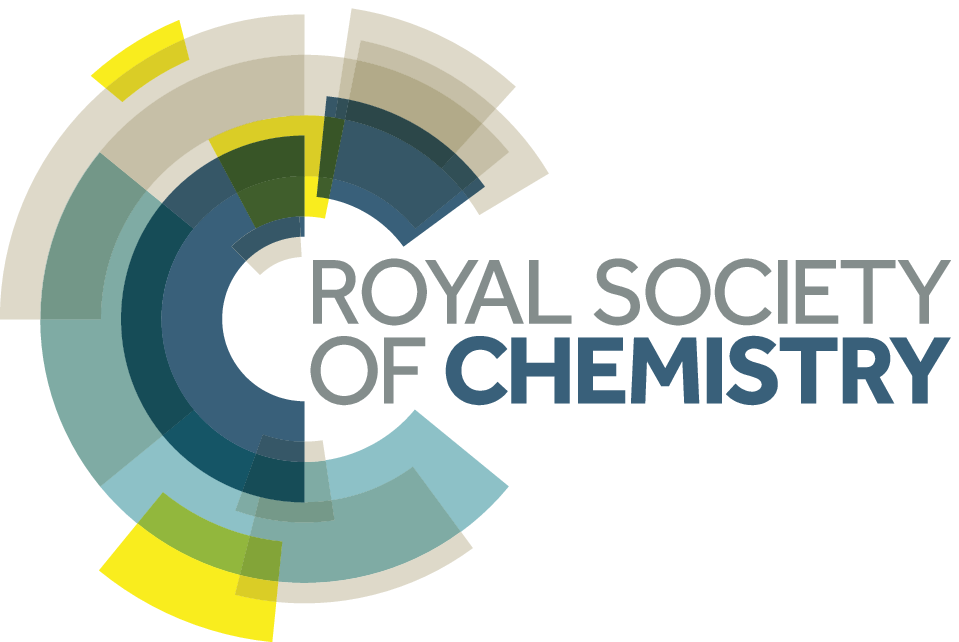}}\\[1ex]
\includegraphics[width=18.5cm]{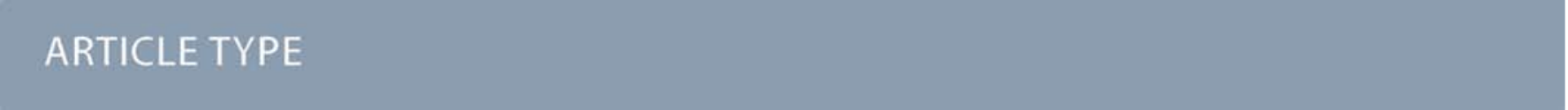}}\par
\vspace{1em}
\sffamily
\begin{tabular}{m{4.5cm} p{13.5cm} }

\includegraphics{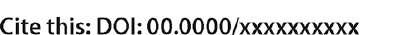} & \noindent\LARGE{\textbf{Synthetic control of structure and conduction properties in Na-Y-Zr-Cl solid electrolytes$^\dag$}} \\
\vspace{0.3cm} & \vspace{0.3cm} \\

 & \noindent\large{Elias Sebti\textit{$^{a,b}$}, Ji Qi\textit{$^{c}$}, Peter M. Richardson\textit{$^{b}$}, Phillip Ridley\textit{$^{c}$}, Erik A. Wu\textit{$^{c}$}, Swastika Banerjee\textit{$^{c}$}, Raynald Giovine\textit{$^{a,b}$}, Ashley Cronk\textit{$^{c}$}, So-Yeon Ham\textit{$^{c}$}, Ying Shirley Meng$^{\ast}$\textit{$^{c,d}$}, Shyue Ping Ong$^{\ast}$\textit{$^{c}$},  and Raphaële J. Clément$^{\ast}$\textit{$^{a,b}$}} \\

\includegraphics{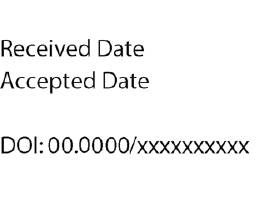} & \noindent\normalsize{In the development of low cost, sustainable, and energy-dense batteries, chloride-based compounds are promising catholyte materials for solid-state batteries owing to their high Na-ion conductivities and oxidative stabilities. The ability to further improve Na-ion conduction, however, requires an understanding of the impact of long-range and local structural features on transport in these systems. In this study, we leverage different synthesis methods to control polymorphism and cation disorder in Na-Y-Zr-Cl solid electrolytes and interrogate the impact on Na-ion conduction. We demonstrate the existence of a more conductive P2$_1$/n polymorph of \ce{Na2ZrCl6} formed upon ball milling. In \ce{Na3YCl6}, the R$\bar{3}$ polymorph is shown to be more conductive than its P2$_1$/n counterpart owing to the presence of intrinsic vacancies and disorder on the Y sublattice. Transition metal ordering in the \ce{Na_{2.25}Y_{0.25}Zr_{0.75}Cl6} composition strongly impacts Na-ion transport, where a greater mixing of \ce{Y^{3+}} and \ce{Zr^{4+}} on the transition metal sublattice facilitates ion migration through partial activation of Cl rotations at relevant temperatures. Overall, Na-ion transport sensitively depends on the phases and transition metal distributions stabilized during synthesis. These results are likely generalizable to other halide compositions and indicate that achieving control over the synthetic protocol and resultant structure is key in the pursuit of improved catholytes for high voltage solid-state sodium-ion batteries.

} \\

\end{tabular}

 \end{@twocolumnfalse} \vspace{0.6cm}

  ]

\renewcommand*\rmdefault{bch}\normalfont\upshape
\rmfamily
\section*{}
\vspace{-1cm}


\footnotetext{\textit{$^{a}$~Materials Department, University of California, Santa Barbara, California 93106, United States}}
\footnotetext{\textit{$^{b}$~Materials Research Laboratory, University of California, Santa Barbara, California 93106, United States}}
\footnotetext{\textit{$^{c}$~Department of NanoEngineering, University of California San Diego, 9500 Gilman Dr., La Jolla, San Diego, California 92093, United States}}
\footnotetext{\textit{$^{d}$~Pritzker School of Molecular Engineering, University of Chicago, Chicago, IL, USA}}

\footnotetext{\dag~Electronic Supplementary Information (ESI) available: [details of any supplementary information available should be included here]. See DOI: 00.0000/00000000.}


\footnotetext{$^{\ast}$Corresponding author emails: shirleymeng@uchicago.edu; ongsp@eng.ucsd.edu; rclement@ucsb.edu}


\section{Introduction}

Solid-state sodium-ion batteries (SSSBs) have emerged over the past few years as an appealing option for safe and scalable renewable energy storage. Owing to abundant global sodium (Na) reserves, SSSBs offer a more affordable and sustainable alternative to lithium-ion (Li-ion) batteries for stationary applications where less stringent gravimetric energy density requirements mean that minimizing raw materials and operating costs is of greater priority\cite{panRoomtemperatureStationarySodiumion2013}. Creating lasting and versatile SSSBs relies on the development of highly conductive, electrochemically and chemically stable solid electrolytes (SEs)\cite{wangDevelopmentSolidstateElectrolytes2019}. Progress towards that end has been made and 10$^{-3}$~S~cm$^{-1}$ conductivities have been achieved with sulfide SEs, on par with liquid electrolytes\cite{hayashiSodiumionSulfideSolid2019, wangInsituInvestigationPressure2018, yuSynthesisUnderstandingNa11Sn2PSe122019}. Nevertheless, these materials still suffer from a narrow electrochemical stability window that makes them incompatible with high voltage cathode materials, inhibiting the development of high voltage SSSBs. 

Since the first report of high Li-ion conductivities in \ce{Li3YCl6} and \ce{Li3YBr6} in 2018 by Asano \textit{et al.}\cite{asanoSolidHalideElectrolytes2018}, halide-based rocksalt-type compounds have gained traction as they simultaneously achieve fast ion transport, good stability at highly oxidative potentials, and are easily processed.
The compositional flexibility of this materials class has led to the synthesis of a wide range of \textit{A}$_3$\textit{MX}$_6$ chemistries (\textit{A} = Li, Na, \textit{M} = Sc, In, Y, Yb, Er, Hf, Ho, Dy, Al, Zr, and \textit{X} = F, Cl, Br, I) and remarkably tunable properties\cite{kwakNa2ZrCl6EnablingHighly2021, kwakNewCostEffectiveHalide2021, schlemLatticeDynamicalApproach2020, schlemNa3XEr12020, liangSiteOccupationTunedSuperionicLixScCl32020, liWaterMediatedSynthesis2019, liOriginSuperionicLi3Y1xInxCl62020, parkHeatTreatmentProtocol2021, parkHighVoltageSuperionic2020, plassEnhancementSuperionicConductivity2022, feinauerUnlockingPotentialFluorideBased2019, asanoSolidHalideElectrolytes2018}. Many of the Li analogues exhibit conductivities on the order of 10$^{-3}$~S~cm$^{-1}$ thanks to a large number of intrinsic cation vacancies and low Li-ion migration barriers due to the weak Coulombic interactions between the Li-ions and the monovalent anion framework\cite{kwakEmergingHalideSuperionic2022, liangSiteOccupationTunedSuperionicLixScCl32020, kwakNewCostEffectiveHalide2021, liOriginSuperionicLi3Y1xInxCl62020}. These conductivities can be further improved through iso- and aliovalent substitution on the halide or metal site\cite{plassEnhancementSuperionicConductivity2022, liuHighIonicConductivity2020, parkHeatTreatmentProtocol2021, helmExploringAliovalentSubstitutions2021, liOriginSuperionicLi3Y1xInxCl62020, schlemNa3XEr12020, wuStableCathodesolidElectrolyte2021, wanInitioStudyDefect2021}. While none of the halides have been shown to be stable against a metal anode, chlorides and fluorides are stable at potentials >4~V as their highly electronegative anions resist oxidation, enabling stable long-term cycling when tested against (uncoated) oxide cathodes\cite{asanoSolidHalideElectrolytes2018}. Finally, scalable solution-based synthesis methods that can yield >100g of material per batch have been reported for many \textit{A}$_3$\textit{MX}$_6$ chemistries\cite{wangUniversalWetchemistrySynthesis2021, liWaterMediatedSynthesis2019}.

The development of highly conductive chloride SEs has been a slow process as their crystal structures, defect concentrations, cation distributions, and resultant ionic conductivities depend sensitively on the synthesis method\cite{stenzelTernareHalogenideVom1993,sebtiStackingFaultsAssist2022, wuStableCathodesolidElectrolyte2021, schlemMechanochemicalSynthesisTool2019, kwakNa2ZrCl6EnablingHighly2021, kwakNewCostEffectiveHalide2021}. In 1993, Stenzel and Meyer first reported the existence of two polymorphs of the Na-conducting \ce{Na3YCl6} (NYC) compound: a low temperature R$\bar{3}$ phase and a high temperature P2$_{1}$/n structure, with a phase transition at 243~K\cite{stenzelTernareHalogenideVom1993}. They synthesized P2$_{1}$/n NYC by slow cooling a melt of the binary precursors, and reported an ionic conductivity of $1 \times10^{-6}$~S~cm$^{-1}$ at 227~$^{\circ}$C.  
Our recent work demonstrated that reducing the material's crystallinity with an additional ball milling step for an NYC sample that had previously been ball milled, annealed at 500~$^{\circ}$C for 24~h, and quenched in a water bath, greatly improved its Na-ion conductivity relative to a sample that had simply been quenched after the 500~$^{\circ}$C annealing step\cite{wuStableCathodesolidElectrolyte2021}. In this study, we also showed that the room temperature conductivity of NYC could be improved by over two orders of magnitude through the substitution of 75\% of the \ce{Y^{3+}} by \ce{Zr^{4+}} and the introduction of cation vacancies. This substitution maintains the P2$_1$/n structure, with the added benefit of incorporating a more Earth-abundant metal (\ce{Zr}) on the \textit{M} site\cite{wuStableCathodesolidElectrolyte2021}. Similar results have been reported for \ce{Na_{3-x}Er_{1-x}Zr_{x}Cl6} by Schlem \textit{et al.}\cite{schlemNa3XEr12020}. Interestingly, the conductivity of \ce{Na_{2.25}Y_{0.25}Zr_{0.75}Cl6} (NYZC75) varies strongly with synthesis and sample processing\cite{wuStableCathodesolidElectrolyte2021}, signaling that the concentration of Na vacancies is not the only predictor of Na-ion transport in this system.   
The other end-member of the \ce{Na_{3-x}Y_{1-x}Zr_{x}Cl6} series, \ce{Na2ZrCl6} (NZC), has also been studied by several groups and conductivities on the order of 10$^{-8}$ and 10$^{-5}$~S~cm$^{-1}$ have been reported when this material was prepared \textit{via} high temperature annealing and \textit{via} high energy ball mill, respectively\cite{wuStableCathodesolidElectrolyte2021, schlemNa3XEr12020, kwakNa2ZrCl6EnablingHighly2021}. 
These results clearly indicate that a better understanding of the impact of synthesis and materials processing on the crystal structure and conduction properties of halide SEs is crucial to their optimization and implementation in SSSBs.

In this study, we investigate the interplay between materials processing, structure, and Na-ion transport in Na-Y-Zr-Cl SEs, paying particular attention to polymorphism and cation disorder. 
In order to tune the structure of NYC, NZC, and NYZC75, we employ ball milling and annealing steps, the latter followed by slow cooling or quenching. We leverage $^{23}$Na solid-state nuclear magnetic resonance (ss-NMR) and theoretical simulations to determine the distribution of Na local environments in these materials and identify the various polymorphic forms of NZC and NYC in samples obtained \textit{via} different synthetic routes.
We report the existence of a high conductivity P2$_1$/n polymorph of NZC generated upon ball milling. The role of polymorphism and cation disorder in facilitating Na-ion transport is probed with variable temperature (VT) NMR, electrochemical impedance spectroscopy (EIS), and bond valence sum mapping. For NYZC75, we combine experimental and computational approaches to trace the origins of synthesis-dependent conduction. For this, we develop an active learning interatomic potential that enables molecular dynamics (MD) simulations of Na-ion transport on large unit cells and over long time scales. Results from these simulations show that the degree of Y-Zr mixing directly impacts the range of Cl rotations, which in turn facilitate Na-ion migration in this class of materials\cite{wuStableCathodesolidElectrolyte2021}. 
In good agreement with these predictions, EIS and VT-NMR indicate that the more disordered, twice ball milled NYZC75 sample is also the most conductive. 
Overall, this study offers important insights into synthesis-dependent conduction in chloride-based SEs, which will guide the future development of highly conductive materials for SSSBs. 

\section{Results and discussion}

\subsection{\ce{Na2ZrCl6}}

\begin{figure}[h]
\centering
  \includegraphics[height=5cm]{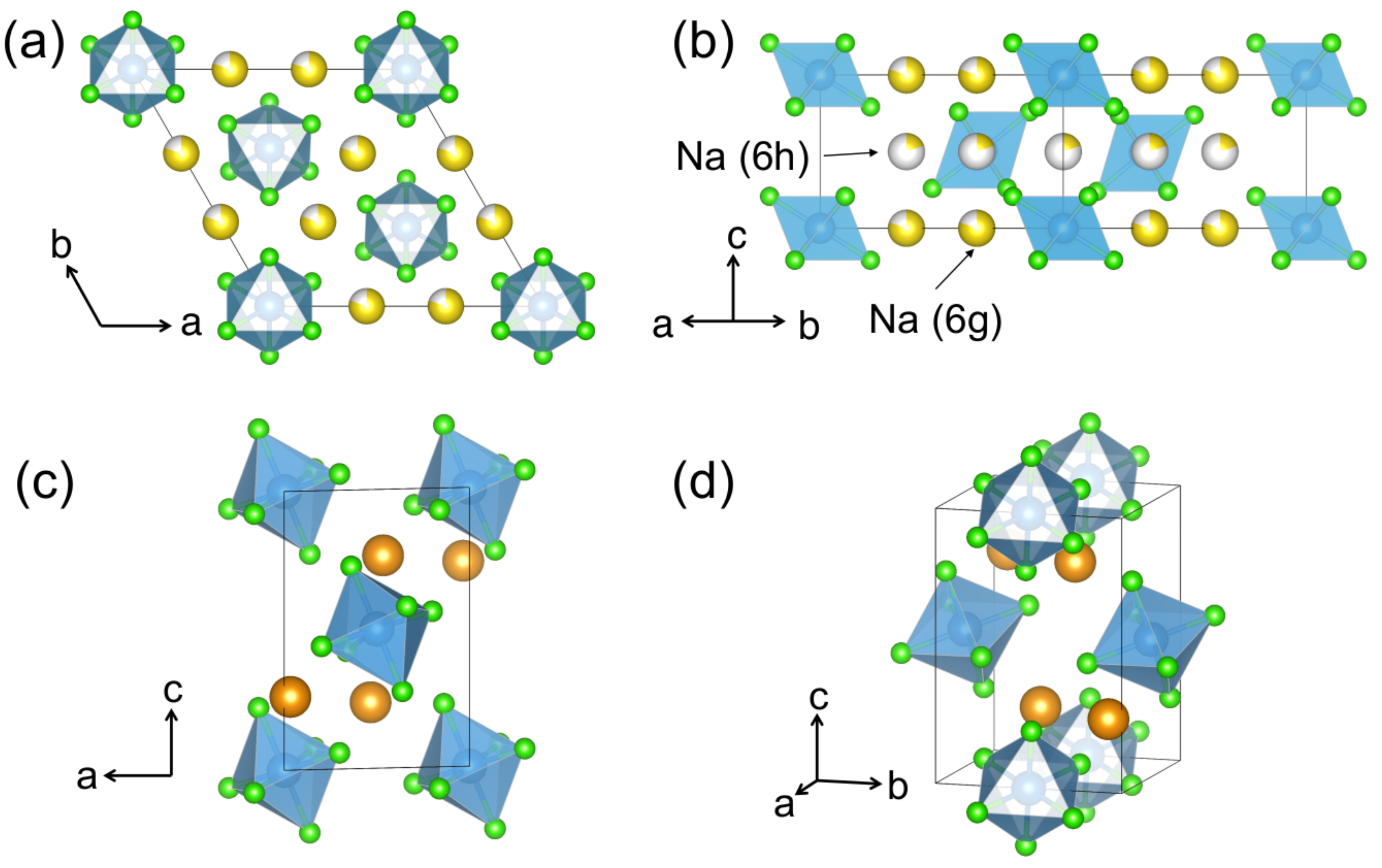}
  \caption{Polymorphs of the \ce{Na2ZrCl6} composition. Color legend: Zr is light blue, Cl is light green, while Na atoms in octahedral and prismatic environments are yellow and orange, respectively. The commonly reported P$\bar{3}$m1 structure as seen (a) down the [$001$] direction and (b) down the [$110$] direction. While some Zr has been reported to populate an extra $2c$ site\cite{wuStableCathodesolidElectrolyte2021}, this site is not shown in this diagram for clarity. P2$_{1}$/n \ce{Na2ZrCl6} is presented in (c-d). (c) When seen down the [010] direction, the body centered arrangement of \ce{ZrCl6^{2-}} octahedra is clearly observable. The arrangement of prismatic Na sites is visible in panel (d).}
  \label{fgr:NZC_structures}
\end{figure}

Hitherto, the crystal structure reported for \ce{Na2ZrCl6} (NZC) has been its P$\bar{3}$m1 form\cite{wuStableCathodesolidElectrolyte2021, kwakNa2ZrCl6EnablingHighly2021}, depicted in  Figure~\ref{fgr:NZC_structures}a-b. This polymorph adopts the \ce{Na2TiF6} archetype structure and closely resembles the better-studied \ce{Li3YCl6} (LYC) SE\cite{bohnsackTernaryChloridesRareEarthe1997, asanoSolidHalideElectrolytes2018}. Layers of octahedral cation sites for \ce{Na+} and \ce{Zr^{4+}} that are edge-sharing in the \textit{ab} plane and face-sharing along the \textit{c}-axis direction form between hexagonally close-packed planes of \ce{Cl-} anions. 
Half of the octahedral sites are vacant -- this is more than in LYC due to the higher valence of \ce{Zr^{4+}} compared to \ce{Y^{3+}}.
Na occupancy is split between the $6g$ and $6h$ Wyckoff positions, while Zr sits in the $1a$ and the $z$ = 0.5 $2d$ sites\cite{wuStableCathodesolidElectrolyte2021, schlemNa3XEr12020}. 
A small fraction of Zr has also been found to occupy an extra $2c$ site, which is analogous to the Y disorder present in ball milled LYC\cite{schlemMechanochemicalSynthesisTool2019, sebtiStackingFaultsAssist2022}.

We also report a P2$_{1}$/n form of NZC, pictured in Figure~\ref{fgr:NZC_structures}c-d. This structure has previously been reported for \ce{Na2IrCl6}\cite{baoNa2IrIVCl6SpinOrbitalInduced2018}, and closely resembles that of the \ce{Na_{3-x}Er_{1-x}Zr_{x}Cl6} and \ce{Na_{3-x}Y_{1-x}Zr_{x}Cl6} compositional series\cite{schlemNa3XEr12020, wuStableCathodesolidElectrolyte2021}. In P2$_{1}$/n NZC, \ce{ZrCl6^{2-}} octahedra are arranged in a body-centered manner with \ce{Zr^{4+}} in the $2a$ sites. Na occupies $4e$ prismatic sites that share corners and edges with one another in the \textit{a}- and \textit{b}-axis directions, respectively. Since NZC contains one less Na per formula unit than NYC, the octahedral interstitial sites that are fully occupied by Na in the NYC structure remain vacant in the NZC structure. 

All NZC samples were prepared with an initial ball milling step followed by an annealing step at 500~$^{\circ}$C for 24~h, and either a water bath quenching step (quenched sample) or a 48~h-long slow cooling step (slow cooled sample). The twice ball milled sample was obtained by ball milling the quenched sample a second time. 
The XRD patterns of the slow cooled, quenched, and twice ball milled samples are compared to the XRD patterns simulated for the P$\bar{3}$m1 and P2$_1$/n forms of NZC in Figure~S1. 
This comparison highlights the presence of P$\bar{3}$m1 NZC in the slow cooled sample, in good agreement with the Le Bail refinement conducted on its XRD pattern and shown in Figure~S2.
For the quenched and twice ball milled samples, the simple pattern comparison points to the coexistence of the P$\bar{3}$m1 and P2$_1$/n forms. In our previous work\cite{wuStableCathodesolidElectrolyte2021}, the XRD pattern collected on quenched NZC was fit with a single P$\bar{3}$m1 phase, suggesting that the P2$_1$/n phase is a very minor component.
For the twice ball milled sample, significant peak broadening precludes an accurate Rietveld refinement of the XRD pattern, although a Le Bail refinement of the diffraction features confirms the presence of both P$\bar{3}$m1 and P2$_1$/n NZC phases (see Figure~S2). 

To better assess the propensity of stabilizing the P$\bar{3}$m1 and P2$_1$/n forms of NZC \textit{via} the three synthesis methods considered here, first principles calculations of phase energetics were conducted using the VASP software package\cite{kresseInitioMolecularDynamics1993, kresseEfficiencyAbinitioTotal1996, kresseEfficientIterativeSchemes1996}. These density functional theory (DFT) calculations indicate that the P$\bar{3}$m1 structure is thermodynamically stable with respect to decomposition into the \ce{NaCl} and \ce{ZrCl4} binary precursors (i.e. it lies on the convex hull). Conversely, the P2$_1$/n structure resides 17~meV atom$^{-1}$ above the hull and is metastable. These results are consistent with the XRD data and suggest that a more equilibrated (e.g. slow cooled) synthesis method results in a sample largely containing the P$\bar{3}$m1 ground state form, while the P2$_1$/n structure can only be obtained through non-equilibrium (e.g. quenched, twice ball milled) synthesis routes.

\begin{figure}[h]
\centering
  \includegraphics[height=6.5cm]{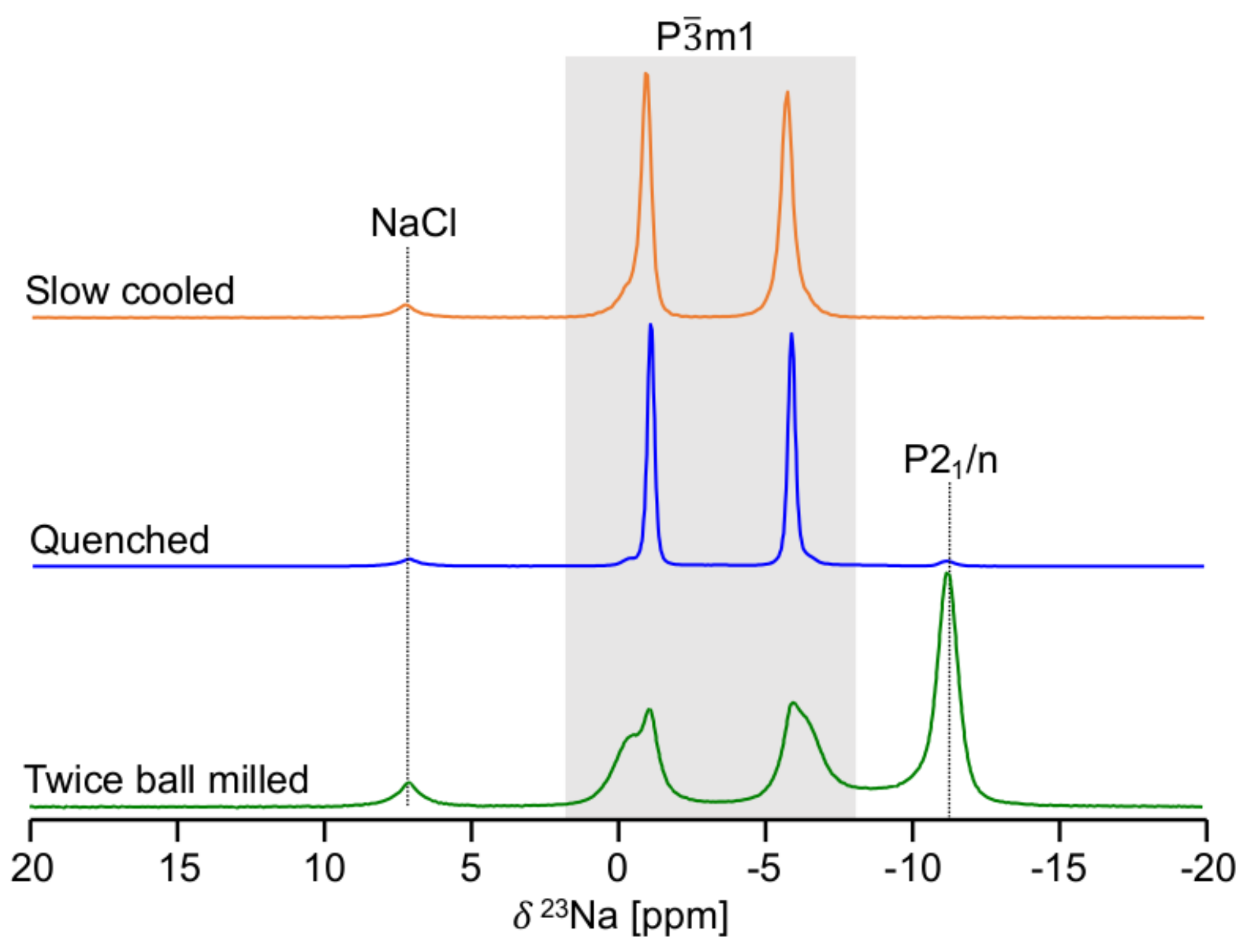}
  \caption{$^{23}$Na ss-NMR spectra for twice ball milled (bottom), quenched (middle), and slow cooled (top) \ce{Na2ZrCl6} samples. Resonances assigned to the P$\bar{3}$m1 polymorph are within the grey box, while the prismatic Na site of the P2$_{1}$/n structure is on the dotted line to its right. Spectra were obtained at a 10 kHz spinning speed, resulting in a sample temperature of 52~$^{\circ}$C.}
  \label{fgr:NZC_NMR}
\end{figure}

Given its sensitivity to crystalline and amorphous phases, $^{23}$Na ss-NMR with magic angle spinning (MAS) was employed to shed further light on the phases formed in the quenched, slow cooled, and twice ball milled NZC samples. The $^{23}$Na ss-NMR spectra are shown in Figure~\ref{fgr:NZC_NMR}. While spectra were collected at room temperature, sample temperatures were around 52~$^{\circ}$C due to frictional heating from the 10~kHz MAS rate. In each spectrum, the 7.2~ppm signal is attributed to an NaCl impurity\cite{wuStableCathodesolidElectrolyte2021}. The spectrum for the quenched sample exhibits two major resonances centered at $-1.1$ and $-6.0$~ppm, each with a small shoulder at $-0.34$ and $-6.6$~ppm, respectively. First principles calculations of NMR parameters using the NMR CASTEP software package\cite{clarkFirstPrinciplesMethods2005, pickardAllelectronMagneticResponse2001, yatesCalculationNMRChemical2007} indicate that the major resonances correspond to Na in the $6g$ and $6h$ octahedral sites of the P$\bar{3}$m1 structure (see Figure~S3). We hypothesize that the low intensity shoulder peaks are associated with Na in $6g$ and $6h$ sites in the vicinity of Zr defect species in $2c$ sites, as evidenced by Rietveld refinement of the XRD pattern obtained on this sample\cite{wuStableCathodesolidElectrolyte2021}. An additional signal is present at $-11$~ppm and is attributed to Na in the $4e$ prismatic sites of the very minor P2$_{1}$/n phase. This signal assignment is corroborated by our NMR CASTEP calculation results shown in Figure~S3. The two main $^{23}$Na resonances and the shoulder peaks observed in the spectrum collected on the slow cooled sample indicate the presence of phase-pure P$\bar{3}$m1 NZC, with a small amount of residual \ce{NaCl}. The slow cooled and quenched samples exhibit rather similar degrees of Zr disorder since the combined integrated intensities of their shoulder peaks account for 11.9 and 14.4\% of the total signal intensity arising from the P$\bar{3}$m1 phase. 
Quenching or slow cooling after a 500~$^{\circ}$C anneal thus strongly favors the formation of the P$\bar{3}$m1 polymorph.
On the other hand, an additional ball milling step facilitates the formation of the P2$_{1}$/n polymorph, as demonstrated by the dominant P2$_{1}$/n resonance in the $^{23}$Na ss-NMR spectrum obtained on the twice ball milled sample.
In this spectrum, the two main $^{23}$Na resonances corresponding to the P$\bar{3}$m1 NZC phase are still present, but a significant increase in the intensity of the shoulder peaks at $-0.34$ and $-6.6$~ppm is observed. As these shoulder peaks are most likely indicative of Zr disorder in the P$\bar{3}$m1 structure, a second ball milling appears to engender greater disorder on the cation lattice, either through the occupation of additional $2c$ sites by Zr, or through the generation of stacking faults as observed in other halide compositions\cite{ sebtiStackingFaultsAssist2022, plassEnhancementSuperionicConductivity2022, liuHighIonicConductivity2020}.
$^{23}$Na EXchange SpectroscopY (EXSY) was conducted at 55~$^{\circ}$C on the twice ball milled sample and provides further evidence for the presence of two NZC phases (see Figure~S4): while Na exchange is clearly observed between the resonances corresponding to the various Na sites in the P$\bar{3}$m1 structure, even at long mixing times of 1~s, no appreciable exchange can be detected with the resonances assigned to Na in the P2$_{1}$/n structure, indicating that the latter Na species are physically separated from the rest of the Na in the P$\bar{3}$m1 phase. 

We now assess the impact of the synthesis method on Na-ion conduction using electrochemical impedance spectroscopy (EIS). We find that the slow cooled sample is least conductive, with an ionic conductivity of 6.6$\times$10$^{-8}$~S~cm$^{-1}$ at room temperature. In our previous work\cite{wuStableCathodesolidElectrolyte2021}, we showed that the slightly higher conductivity of the quenched sample ($1.4\times10^{-7}$~S cm$^{-1}$) could be increased by over two orders of magnitude by ball milling a second time, resulting in a conductivity of $2.6\times10^{-5}$~S cm$^{-1}$ at room temperature.
These results agree well with work from Kwak \textit{et al.} that showed that the conductivity of a ball milled NZC sample dropped from $1.8\times10^{-5}$~S cm$^{-1}$ to $6.9\times10^{-8}$~S cm$^{-1}$ after a 12~h annealing step at 400~$^{\circ}$C\cite{kwakNa2ZrCl6EnablingHighly2021}. Given that both the fraction of the P2$_{1}$/n phase and cation disorder within the P$\bar{3}$m1 phase sharply increase after a second ball milling step (see Figure 2), it is not possible to trace back the increased Na conductivity to a single structural feature. Further, while Kwak \textit{et al.}\cite{kwakNa2ZrCl6EnablingHighly2021} attributed these conductivity improvements to increased disorder within the P$\bar{3}$m1 phase, they did not conduct $^{23}$Na ss-NMR experiments and it therefore remains unclear whether their low crystallinity ball milled NZC sample contained a P2$_{1}$/n phase that could instead be responsible for fast Na-ion transport. 

\begin{figure}[h]
\centering
  \includegraphics[height=4cm]{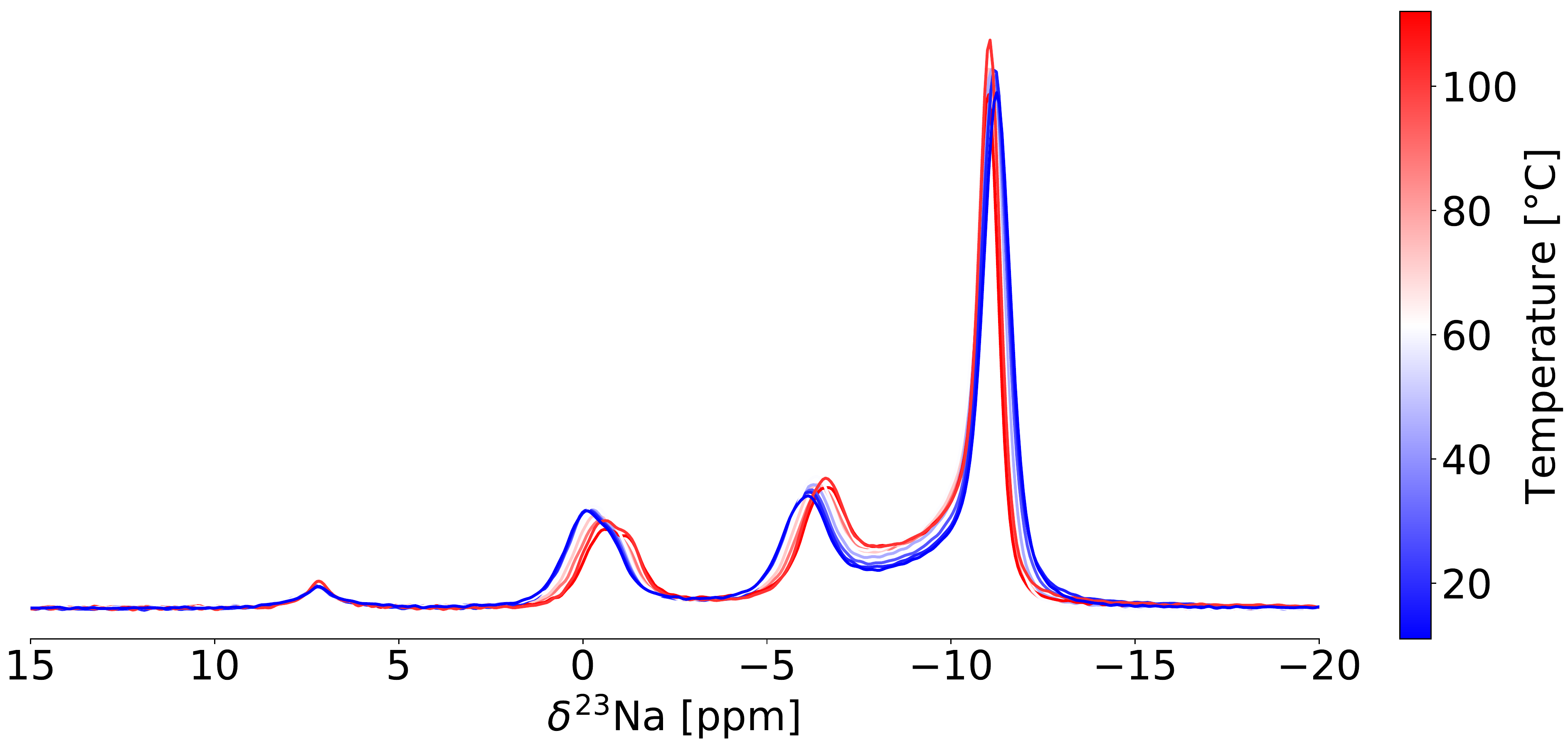}
  \caption{Variable temperature $^{23}$Na ss-NMR on twice ball milled {\ce{Na2ZrCl6}}. Spectra are color-coded using the temperature scale on the right of the plot. The peak at 7.2 ppm is NaCl. All spectra were obtained at 18.8~T with a 10 kHz MAS speed.}
  \label{fgr:NZC_VT-NMR}
\end{figure}

Variable temperature (VT) $^{23}$Na ss-NMR was conducted between 11~$^{\circ}$C and 112~$^{\circ}$C on the twice ball milled NZC sample to identify the origin of the Na-ion conductivity improvements after a second ball milling step. In NMR, as the sample temperature is raised and the rate of chemical exchange between distinct local environments increases, the resonances corresponding to the sites in exchange grow slightly closer together, broaden, and eventually coalesce at their weighted average position\cite{levittSpinDynamicsBasics2008}. As the temperature is further increased, those average resonances sharpen through a process known as motional narrowing. As such, VT-NMR is a useful tool to determine the local environments participating in Na-ion motion, and can thus identify which polymorph is more conductive in a phase mixture; this type of information is inaccessible with bulk conductivity techniques such as EIS.
An example fit of the $^{23}$Na ss-NMR spectrum collected at 45~$^{\circ}$C is presented in Figure~S5, and plots showing the temperature evolution of the resonant frequencies and linewidths of the $^{23}$Na signals corresponding to the different Na local environments in the P$\bar{3}$m1 and P2$_1$/n phases are shown in Figure~S6. Because the validity of the VT-NMR approach depends on the stability of our samples during the measurements, we note that no phase transition nor any change in the phase fraction of the P$\bar{3}$m1 and P2$_1$/n polymorphs was detected based on fits of the $^{23}$Na ss-NMR spectra over the entire temperature range probed here. Our results clearly indicate that the P2$_1$/n resonance undergoes motional narrowing, suggesting more frequent Na-ion hops between prismatic sites in this structure as the temperature increases from 11 to 112~$^{\circ}$C. Conversely, the linewidths of the P$\bar{3}$m1 resonances do not evolve monotonically with temperature and do not show any sign of coalescence. These results indicate that the P2$_1$/n form is most conductive.
Over the probed temperature range, the resonances corresponding to Na sites in the P$\bar{3}$m1 polymorph and to prismatic Na in the P2$_1$/n polymorph grow closer together, indicating a small degree of Na exchange between the two phases that could not be detected in the EXSY experiment at 55~$^{\circ}$C. These findings are corroborated by another set of VT-NMR spectra obtained on the quenched NZC (see Figure~S7), indicating a small amount of Na exchange between the P$\bar{3}$m1 and P2$_1$/n phases without any observable narrowing or coalescence of the P$\bar{3}$m1 resonances. 

\subsection{\ce{Na3YCl6}}

\begin{figure}[h]
\centering
  \includegraphics[height=4cm]{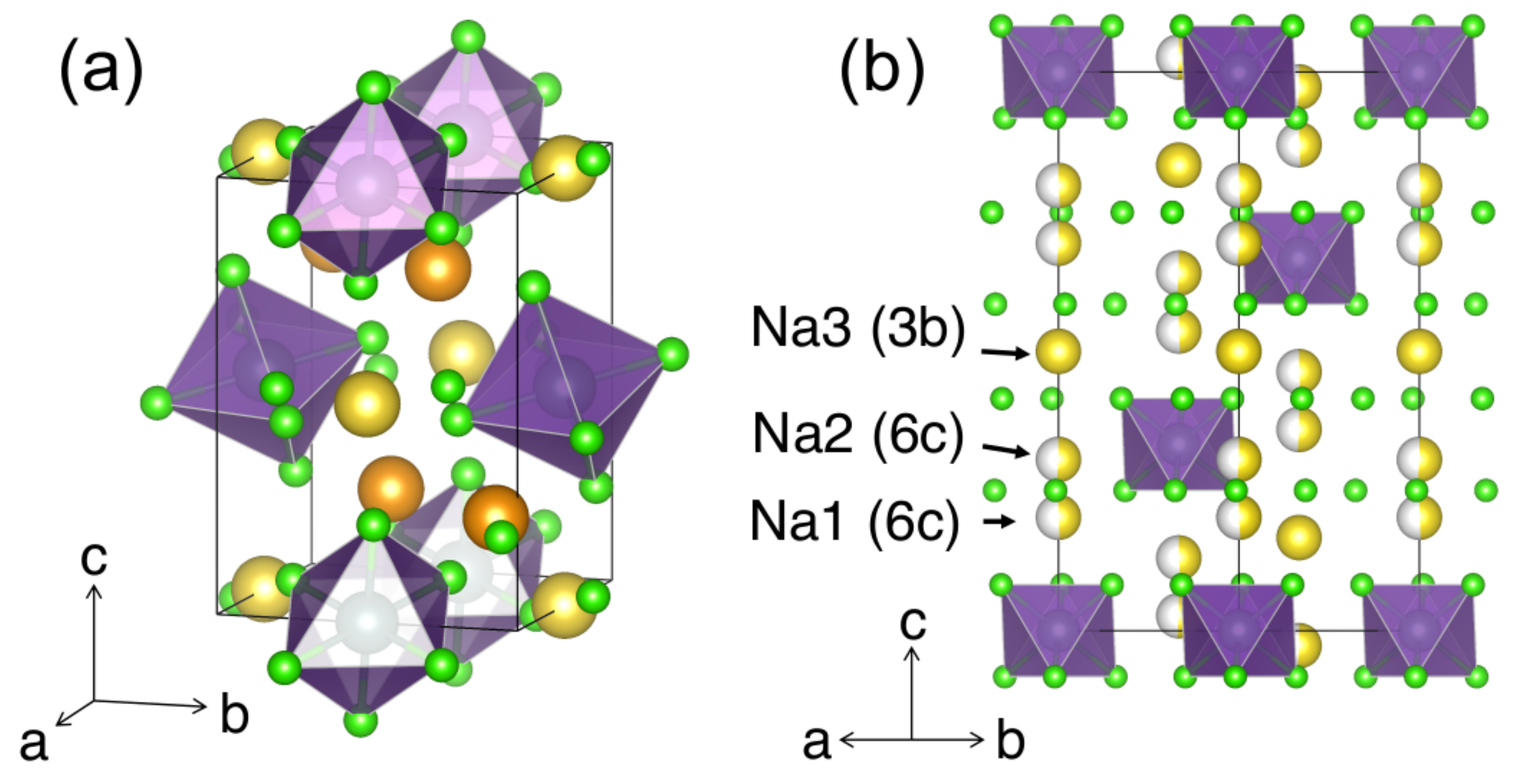}
  \caption{Polymorphs of the \ce{Na3YCl6} composition. Color legend: Y is purple, Cl is light green, while Na atoms in octahedral and prismatic environments are yellow and orange, respectively. (a) The P2$_{1}$/n structure with Na occupying prismatic and octahedral sites. (b) The R$\bar{3}$ polymorph, as seen down the [110] direction. The three Na sites are depicted with varying degrees of occupancy, as reported by Stenzel and Meyer{\cite{stenzelTernareHalogenideVom1993}}.}
  \label{fgr:NYC_structures}
\end{figure}

Similar to NZC, \ce{Na3YCl6} (NYC) experiences polymorphism and can adopt a structure with either R$\bar{3}$ or P2$_{1}$/n symmetry. Both structures have been reported \cite{stenzelTernareHalogenideVom1993, wuStableCathodesolidElectrolyte2021} and representations of each are given in Figure~\ref{fgr:NYC_structures}. The P2$_{1}$/n structure is a cryolite-type structure with a prismatic (2x multiplicity) and an octahedral Na site. These prismatic and octahedral Na sites share edges with one another, while the \ce{YCl6^{3-}} octahedra are arranged in a body-centered manner. The R$\bar{3}$ form has been referred to as a stuffed \ce{LiSbF6}-type structure, where Na and Y species occupy octahedral sites\cite{stenzelTernareHalogenideVom1993}. These octahedral cation sites share edges with one another in the \textit{ab} plane, and form face-sharing chains of five \ce{NaCl6^{5-}} octahedra followed by one \ce{YCl6^{3-}} octahedron along the \textit{c}-axis direction. The structure contains three distinct Na sites (at $z$ = 0.2 $6c$, $z$ = 0.3 $6c$, and $z$ = 0.5 $3b$) and a third of its cation sites are empty.

Slow cooled, quenched, and twice ball milled NYC samples were prepared in an analogous manner to the NZC samples discussed earlier.
The XRD patterns collected on the three NYC samples are shown in Figure~S8 and compared to the simulated XRD patterns for the R$\bar{3}$ and P2$_1$/n NYC polymorphs reported in the ICSD\cite{wuStableCathodesolidElectrolyte2021, stenzelTernareHalogenideVom1993}. The data indicate that slow cooled NYC consists entirely of the R$\bar{3}$ polymorph, while the quenched and twice ball milled samples are composed of a mixture of the R$\bar{3}$ and P2$_1$/n phases. While we attempted a Rietveld refinement of the diffraction pattern for our slow cooled NYC, Y fluorescence precluded detailed analysis of the diffraction pattern. 
The broad diffraction peaks of the twice ball milled sample reflect its low crystallinity, as observed in other ball milled halide systems\cite{kwakNa2ZrCl6EnablingHighly2021, wuStableCathodesolidElectrolyte2021, asanoSolidHalideElectrolytes2018}. Scanning electron microscopy (SEM) images for each sample, shown in Figure~S9, demonstrate that the twice ball milled sample is composed of the smallest particles (1-20~$\mu$m), while the largest particles (10-40~$\mu$m) are observed for the slow cooled sample.

\begin{figure}[h]
\centering
  \includegraphics[height=6cm]{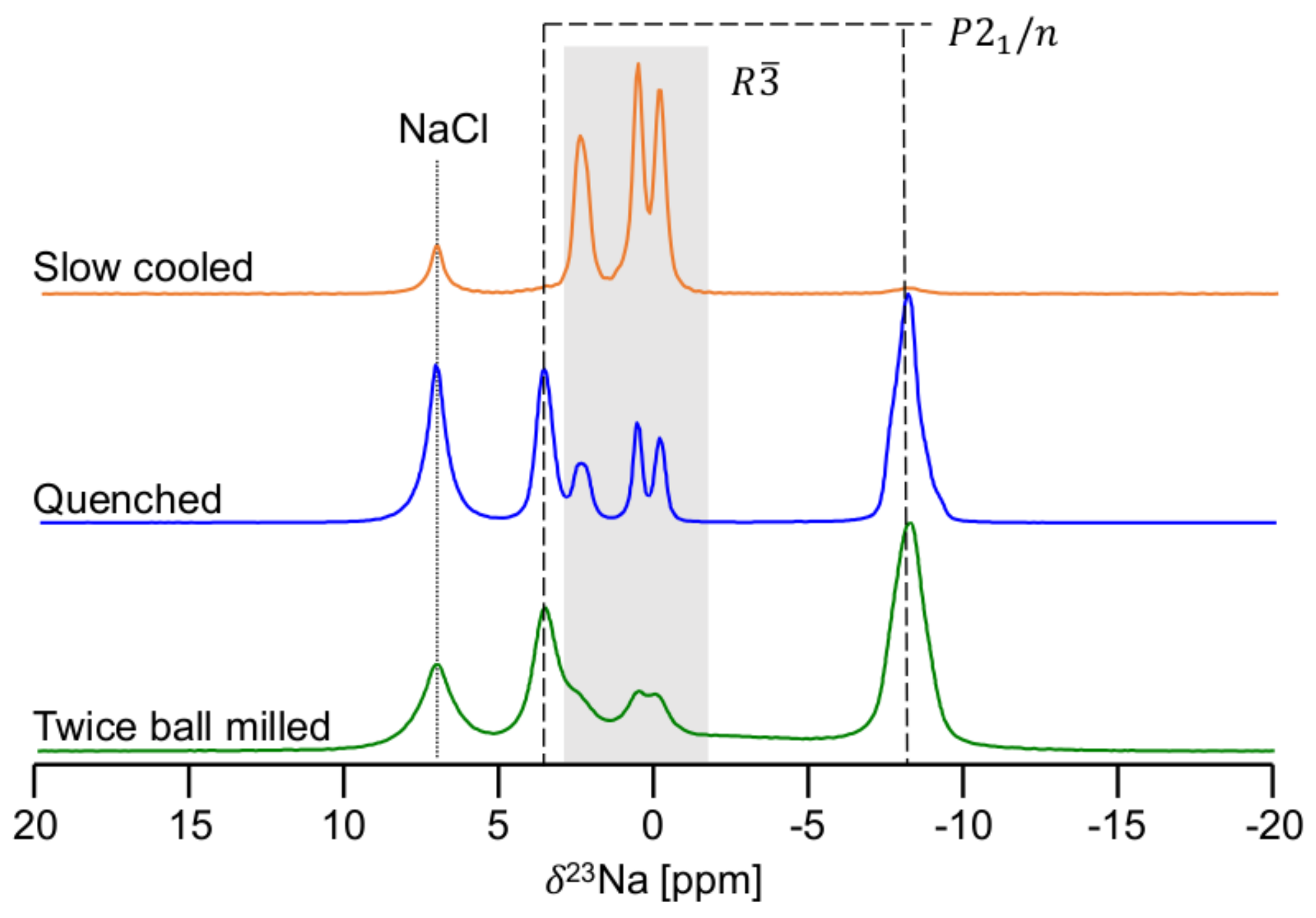}
  \caption{$^{23}$Na ss-NMR spectra for twice ball milled (bottom), quenched (middle), and slow cooled (top) \ce{Na3YCl6} samples. Resonances assigned to the octahedral sites of the R$\bar{3}$ polymorph are within the grey box, while the octahedral and prismatic Na sites of the P2$_{1}$/n structure are indicated by the dashed lines. Spectra with a 10~kHz MAS speed, resulting in a sample temperature of 52~$^{\circ}$C. The spectrum for the twice ball milled sample is replicated from our previous work\cite{wuStableCathodesolidElectrolyte2021} for easier comparison with the other two samples.}
  \label{fgr:NYC_NMR_var_prep}
\end{figure}

$^{23}$Na ss-NMR was employed to determine the local Na environments present in the slow cooled, quenched, and twice ball milled samples (see Figure~\ref{fgr:NYC_NMR_var_prep}). All spectra exhibit a peak at 7.2 ppm corresponding to leftover NaCl precursor. The spectrum obtained on the slow cooled sample is comprised of three prominent resonances centered at 2.6, 0.7, and $0.0$~ppm, presumably corresponding to the three octahedral Na sites in the R$\bar{3}$ structure based on the XRD results discussed earlier. 
The signals are present in a 1:1:1 ratio, which is consistent with half occupation of the $6c$ sites and full occupation of the $3b$ sites in the R$\bar{3}$ NYC structure reported by Stenzel and Meyer\cite{stenzelTernareHalogenideVom1993}. The three R$\bar{3}$ signals are also present in the spectra collected on the quenched and twice ball milled samples, although the spectra are dominated by resonances at 3.6 and $-8.1$~ppm that presumably correspond to Na sites in the second P2$_{1}$/n phase observed by XRD. Those signal assignments are corroborated by NMR CASTEP calculations that predict that the $^{23}$Na resonances corresponding to Na sites in the R$\bar{3}$ phase are flanked by those associated with Na sites in P2$_1$/n structure, as shown in Figure~S10. The NMR CASTEP calculations further enable the assignment of the 3.6 and $-8.1$~ppm resonances to Na in octahedral and prismatic sites in the P2$_1$/n structure, respectively, while $^{23}$Na EXSY at 55~$^{\circ}$C on the twice ball milled sample confirms the presence of two separate phases (R$\bar{3}$ and P2$_1$/n) with no detectable chemical exchange between them (see Figure~S11). This in-depth analysis of the slow cooled, quenched, and twice ball milled samples goes significantly beyond our previous work, where we mistakenly assigned the resonances associated with the R$\bar{3}$ phase in the $^{23}$Na ss-NMR spectrum collected on the twice ball milled sample to disorder within the P2$_1$/n phase\cite{wuStableCathodesolidElectrolyte2021}. Interestingly, the $^{23}$Na ss-NMR data indicate that NaCl and the R$\bar{3}$ and P2$_1$/n forms of NYC are present in similar amounts in the quenched and twice ball milled samples (25.9$\%$ \ce{NaCl}, 19.5$\%$ R$\bar{3}$, and 54.6$\%$ P2$_{1}$/n), suggesting that, unlike in NZC, the second ball milling step does not dictate the phase composition of samples quenched after the 500~$^{\circ}$C annealing step. As expected, the $^{23}$Na ss-NMR resonances are significantly broader for the low crystallinity twice ball milled sample than for the more crystalline quenched sample. Notably, the R$\bar{3}$ and P2$_{1}$/n phase fractions in the three NYC samples observed by $^{23}$Na ss-NMR are consistent with the relative stability of the two polymorphs obtained from first principles VASP calculations. The thermodynamically-equilibrated slow cooled sample consists of a single R$\bar{3}$ phase that is predicted to reside on the convex hull, while the additional P2$_1$/n phase present in the quenched and twice ball milled samples is found to be only slightly metastable, lying 8.8~meV~atom$^{-1}$ above the convex hull.

EIS measurements on the twice ball milled and quenched samples yield ionic conductivities of $9.5\times10^{-8}$~S cm$^{-1}$ at 25~$^{\circ}$C and $1.6\times10^{-11}$~S cm$^{-1}$ at 60~$^{\circ}$C, respectively\cite{wuStableCathodesolidElectrolyte2021}. The ionic conductivity of the slow cooled sample was too low to be measured at 60~$^{\circ}$C, suggesting that it is even lower than that of the quenched sample. While $^{23}$Na VT-NMR was attempted on the quenched NYC sample to elucidate the relative conduction properties of the R$\bar{3}$ and P2$_1$/n forms of NYC, much of the P2$_{1}$/n phase converted into the R$\bar{3}$ phase over the course of the measurement, preventing a reliable analysis the results (see Figure~S12). Room temperature spectra (corresponding to a sample temperature of 52~$^{\circ}$C) taken before and after the VT-NMR measurements confirm that the observed structural changes are permanent (see Figure~S13). $^{23}$Na VT-NMR measurements on the twice ball milled sample were not attempted as the same phase transition is expected to occur.

\subsection{Synthetic Control of Polymorphism and Na-ion Transport in \ce{Na2ZrCl6} and \ce{Na3YCl6}}

The results presented in the last two sections revealed some of the links between synthesis conditions, phase composition of the resulting NZC and NYC samples, and Na-ion conduction properties. Here, we provide further insight into the microscopic mechanisms underlying Na-ion motion through the various structures discussed previously, and relate phase stability with phase formation \textit{via} the various synthetic approaches explored in this work.
In the P2$_{1}$/n structure shared by NYC and NZC, Na-ion transport is isotropic and occurs \textit{via} hops between edge-sharing prismatic and octahedral sites, as illustrated by the bond valence sum maps of Figure S14. In NZC, the occupied prismatic (Prism) Na sites share edges with the vacant octahedral (Oct) sites, creating 3D Prism-Oct-Prism pathways for rapid Na-ion transport. NYC, on the other hand, has one more Na per formula unit and all of the available Na sites are filled, drastically limiting Na-ion motion.  

The layered R$\bar{3}$ and P$\bar{3}$m1 forms of NYC and NZC share similar anisotropic Na-ion hopping pathways, also depicted in the bond valence sum maps of Figure~S14. 
In the \textit{ab} plane, Na-ions migrate between edge-sharing octahedral sites \textit{via} bridging tetrahedral (Tet) interstitial sites (Oct-Tet-Oct). Neighboring \ce{NaCl6^{5-}} octahedra share a face in the \textit{c}-axis direction, leading to a direct Oct-Oct pathway for Na-ion migration normal to the \textit{ab} planes. Notably, intrinsic vacancies on the $6c$ sites of the R$\bar{3}$ form of NYC and on the $6g$ and $6h$ sites of the P$\bar{3}$m1 form of NZC provide empty sites for Na-ion migration. 
The bond valence sum map of the R$\bar{3}$ NYC structure reflects the facile out-of-plane Na-ion Oct-Oct hopping between face-sharing octahedral Na sites and worse Oct-Tet-Oct in-plane transport, in good agreement with a recent computational study\cite{qieYttriumSodiumHalides2020}. Connectivity chains between \ce{NaCl6^{5-}} octahedra along the \textit{c}-axis are punctuated by \ce{YCl6^{3-}} octahedra. Disorder on the Y sublattice induced by ball milling, as observed in \ce{Li3YCl6}, could be responsible for enhancing the connectivity between Na sites and improving Na-ion conduction\cite{sebtiStackingFaultsAssist2022, schlemMechanochemicalSynthesisTool2019}.

The enhanced conductivity of the twice ball milled NZC sample is explained by the superior Na-ion transport properties of the P2$_1$/n NZC phase. While both polymorphs of NZC contain intrinsic vacancies, the isotropic Na-ion hopping pathways of the P2$_1$/n form further enhance conductivity relative to the anisotropic pathways of the P$\bar{3}$m1, likely because conduction does not depend on the alignment of neighboring grains.    
Given its lack of intrinsic vacancies, the NYC P2$_1$/n form is not expected to be very conductive despite its emergence in the more conductive quenched and twice ball milled samples. 
Instead, we expect that the lower crystallinity and/or increased cation disorder within the R$\bar{3}$ phase is responsible for the enhanced conductivity of the twice ball milled sample, as observed for \ce{Li3YCl6}\cite{asanoSolidHalideElectrolytes2018, schlemMechanochemicalSynthesisTool2019, sebtiStackingFaultsAssist2022}. 

Based on the conduction mechanisms discussed above, the design of high conductivity chloride SEs should first focus on the inclusion of intrinsic cation vacancies, and then on the presence of isotropic conduction pathways. 
Those two conditions can be satisfied through aliovalent doping and concurrent introduction of Na vacant sites into a structure with an isotropic conduction mechanism. This strategy is illustrated here with \ce{Na_{2.25}Y_{0.25}Zr_{0.75}Cl6}, discussed in the next section.

The results presented here constitute an important exploration of the links between synthesis, polymorphism, and ion transport properties in a family of Na-ion conductors.
For NZC, ball milling is required to generate appreciable amounts of the higher energy P2$_1$/n phase, which is consistent with its relatively large energy above the hull (17 ~meV~atom$^{-1}$). $^{23}$Na VT-NMR indicates that the P2$_1$/n structure is more conductive than the P$\bar{3}$m1 structure, even with the introduction of greater Zr site disorder in the P$\bar{3}$m1 phase (as evidenced by the growth of the shoulder peaks in the $^{23}$Na NMR spectrum). Yet, the P2$_1$/n phase fraction is only of about 43\% in the twice ball milled sample, based on $^{23}$Na NMR signal integration, and increasing this phase fraction could yield further improvements in bulk conductivity. Faster rotational speeds and longer milling times could offer a pathway towards this end. For NYC, we find that slow cooling from 500~$^{\circ}$C to room temperature over the course of 48~h allows for the isolation of the R$\bar{3}$ phase. Interestingly, Stenzel and Meyer obtained a single P2$_1$/n phase when slow cooling from the melt, suggesting that the exact parameters used for the slow cool (not reported in their study) determine the phases formed\cite{stenzelTernareHalogenideVom1993}. These findings are consistent with the small (8.8~meV~atom$^{-1}$) energy difference between the two polymorphs, as calculated by DFT. While the R$\bar{3}$ and P2$_1$/n phase fractions in the quenched and twice ball milled NYC samples are similar, the conductivities diverge by three orders of magnitude. One important difference between the two samples is the crystallinity and transition metal ordering within the two phases, as observed \textit{via} XRD and NMR. Additional exploration of the influence of crystallinity on ionic conduction in halides is of interest and will be the subject of future work. In their investigation of the \ce{Ag_{1-x}Na_xYCl6} system, Stenzel and Meyer observed an increase in the R$\bar{3}$ to P2$_1$/n phase transition temperature at higher \ce{Ag} contents, suggesting that the phase stability is sensitive to the composition\cite{stenzelTernareHalogenideVom1993}. Exploring the influence of substituting Y by isovalent ions with different ionic radii on NYC phase behavior could enable the stabilization of higher phase fractions of the R$\bar{3}$ phase, the introduction of additional disorder on the Y sublattice, and improvements in overall sample conductivity.


\subsection{\ce{Na_{2.25}Y_{0.25}Zr_{0.75}Cl6}}
\begin{figure}[H]
\centering
  \includegraphics[height=4.7cm]{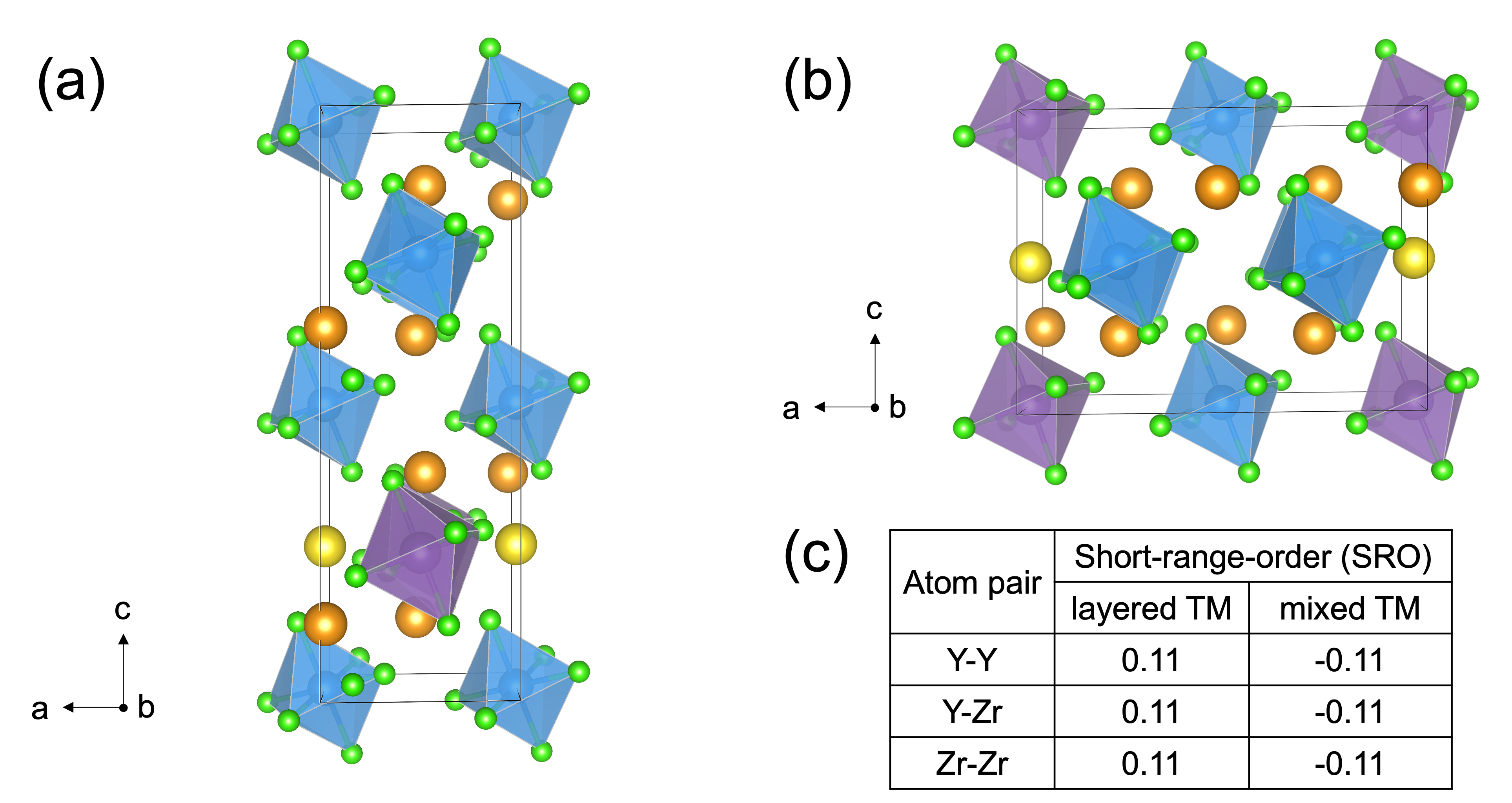}
  \caption{\ce{Na_{2.25}Y_{0.25}Zr_{0.75}Cl6} structural models with different Y-Zr orderings. Color legend: Y is purple, Zr is light blue, Cl is light green, while Na in an octahedral or prismatic environment is yellow and orange, respectively. The two types of Y-Zr orderings are referred to as (a) layered-TM and (b) mixed-TM structures. (c) Short-range order (SRO) parameters that describe Y-Y, Y-Zr and Zr-Zr correlations. These SRO parameters are defined in the Methods section.}
  \label{fgr:NYZC75_structures_SRO}
\end{figure}

Slow cooled, quenched, and twice ball milled \ce{Na_{2.25}Y_{0.25}Zr_{0.75}Cl6} (NYZC75) samples were prepared in an analogous manner to NZC and NYC samples discussed earlier. A Le Bail refinement of the XRD pattern obtained on the slow cooled sample indicates that the material is highly crystalline and exhibits a P2$_1$/n symmetry, as shown in Figure~S15. In our previous work, we showed that quenched and twice ball milled NYZC75 also adopt the P2$_{1}$/n crystal structure\cite{wuStableCathodesolidElectrolyte2021}. We note that while the structure of the low crystallinity, twice ball milled sample is difficult to ascertain from XRD alone, further evidence for a shared structure with the quenched sample comes from Raman spectra indicating similar \ce{ZrCl6^{2-}} connectivities (see Figure~S16).

\begin{figure}[H]
\centering
  \includegraphics[height=13cm]{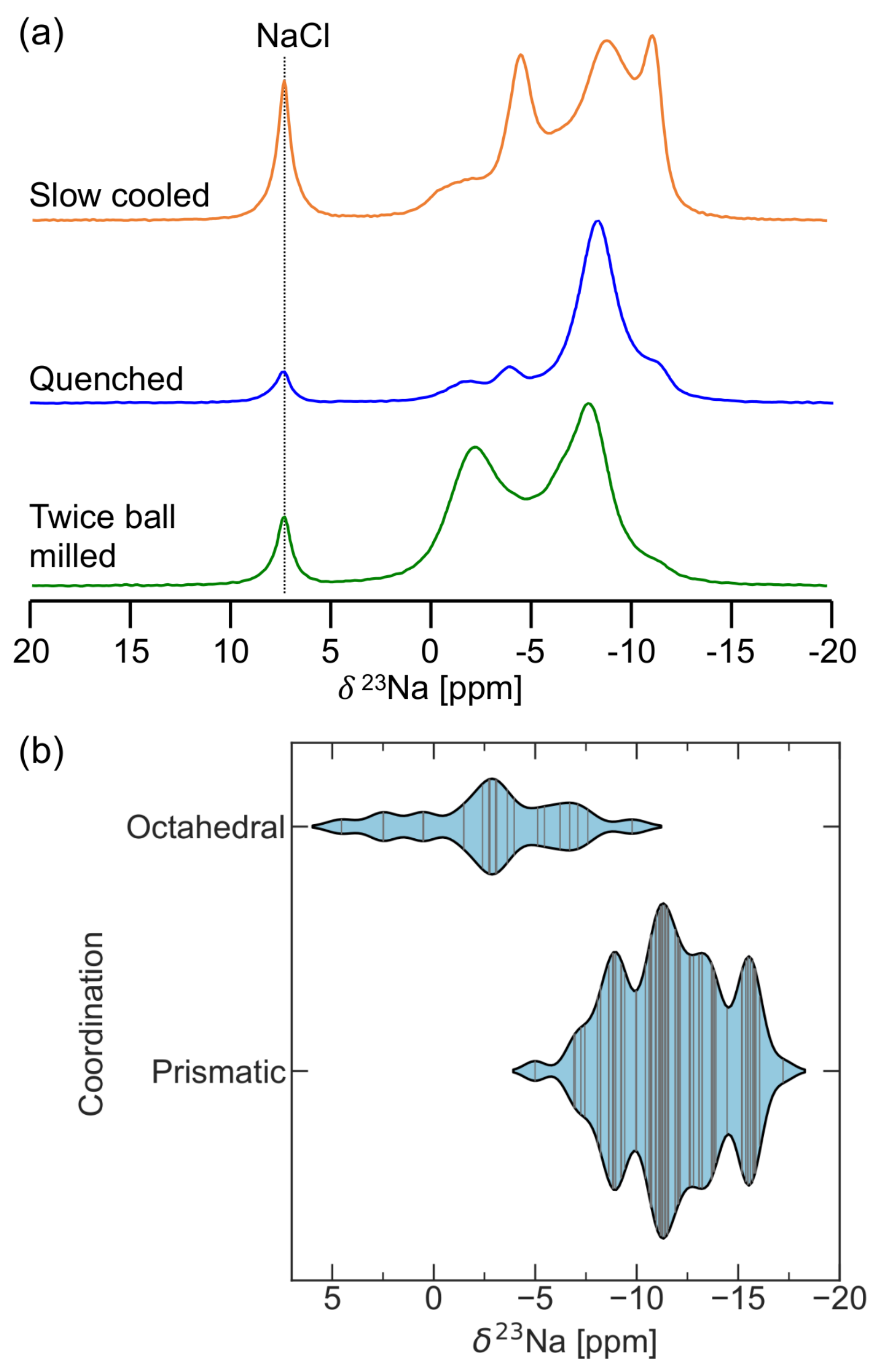}
  \caption{(a) $^{23}$Na ss-NMR spectra for twice ball milled (bottom), quenched (middle), and slow cooled (top) \ce{Na_{2.25}Y_{0.25}Zr_{0.75}Cl6}. Leftover \ce{NaCl} precursor leads to a characteristic signal at 7.2~ppm. Spectra were obtained at 52~$^{\circ}$C using a 10~kHz spinning speed. (b) Violin plots showing the distribution of $^{23}$Na isotropic shift values ($\delta^{23}$Na) computed with CASTEP for octahedral and prismatic Na sites in select \ce{Na_{2.25}Y_{0.25}Zr_{0.75}Cl6} supercells. }
  \label{fgr:NYZC75_NMR&CASTEP}
\end{figure}

To gain insight into the propensity for disorder on the transition metal and Na sublattices of NYZC75, we enumerated Na-vacancy and Y-Zr orderings within a model P2$_1$/n structure and computed their energies using DFT calculations. A total of 229 supercells were generated based on two distinct Y-Zr transition metal (TM) orderings, referred to as ``layered-TM'' and ``mixed-TM'' structures (see Figure~\ref{fgr:NYZC75_structures_SRO}a-b). The layered-TM structures consist of one layer of \ce{YCl6^{3-}} octahedra alternating with three layers of \ce{ZrCl6^{2-}} octahedra. In contrast, the mixed-TM structures consist of one layer of \ce{ZrCl6^{2-}} octahedra followed by a layer containing equal numbers of \ce{ZrCl6^{2-}} and \ce{YCl6^{3-}} octahedra. Figure~\ref{fgr:NYZC75_structures_SRO}c shows the Warren-Crowley chemical short-range order (SRO) parameters between Y and Zr species in these structures (see Methods section). The positive SRO parameters in layered-TM structures indicate attraction between like species (Y-Y and Zr-Zr) and repulsion between dissimilar species (Y-Zr), leading to the layered configuration. The opposite is true for the mixed-TM structures. Of the 12 Na sites in the P2$_1$/n NYZC75 supercells considered (4 octahedral and 8 prismatic), only 9 are occupied. To better understand the impact of octahedral and prismatic site occupation by Na on energetics, the octahedral Na site occupancy ($O^{Na}_{octahedral}$) was varied from 25\% to 100\% in 25\% intervals. As shown in Figures~S17 and S18, the lowest energy Na-vacancy ordering maximizes the prismatic site occupancy and minimizes octahedral Na site occupancy for both the layered-TM and mixed-TM structures, although octahedral Na site occupation only leads to a small energy penalty ($\sim 5-25$ meV~atom$^{-1}$). Furthermore, the lowest TM ordering is the layered arrangement, but again the mixed ordering is only slightly higher in energy ($\sim 5-10$ meV~atom$^{-1}$). In summary, these results indicate that the ground state structure is an ordered one with a layered configuration of TM species and Na fully occupying the prismatic sites and 25\% of the octahedral sites, but also that TM disorder and Na-vac disorder are readily accessible via typical synthetic levers such as temperature and ball milling.

$^{23}$Na ss-NMR was employed to gain insight into the Na local environments present in the three NYZC75 samples. The spectra are shown in Figure~\ref{fgr:NYZC75_NMR&CASTEP}a. NaCl left over from the synthesis is once again observed at 7.2~ppm. Three broad $^{23}$Na resonances feature prominently in the spectrum collected on the twice ball milled compound. These resonances are composed of many overlapping signals, indicating a wide distribution of Na sites in the sample. EXSY experiments at 55~$^{\circ}$C reveal that, aside from NaCl, all of the broad resonances are in exchange with one another, indicating the presence of a single NYZC75 phase (see Figure~S19). The $^{23}$Na resonances in the spectra collected on the quenched and slow cooled samples are better resolved. The improved resolution is likely due to the higher crystallinity of these samples and a more ordered Y-Zr sublattice, leading to fewer Na local environments. 

To facilitate the interpretation of the $^{23}$Na NMR spectra, NMR CASTEP calculations were carried out on select NYZC75 supercells. Overall, the computed $\delta$($^{23}$Na) values fall within the ppm range of the experimentally observed resonances, indicating that our enumerated orderings are representative of the actual structure. 
As seen in Figure~\ref{fgr:NYZC75_NMR&CASTEP}b, $\delta$($^{23}$Na) values $\geq-5$~ppm correspond to Na in octahedral sites, while Na in prismatic sites exhibit more negative shifts in the $-5$ to $-17.5$~ppm range. The spectrum obtained for the twice ball milled sample displays the most intensity in the 0 to $-5$~ppm range, suggesting that this compound has the greatest octahedral Na occupancy of all three samples examined here. Low octahedral occupancy in the quenched sample is in line with our DFT calculations and the Rietveld refinements conducted by Schlem \textit{et al.}\cite{schlemNa3XEr12020} on annealed \ce{Na_{3-x}Er_{1-x}Zr_xCl6} samples, which is isostructural to NYZC. With increasing $x$ values, these authors showed that the prismatic sites remain fully occupied, while the octahedral site occupancy drops. Here, high energy ball milling appears to stabilize a greater distribution of Na local environments (including higher energy ones) and therefore, more Na residing in octahedral sites. 
Slow cooling, on the other hand, leads to significant intensity at the most negative resonance around $-13$~ppm. NMR CASTEP calculations reveal that, for Na in prismatic sites, the presence of a greater number of Zr next nearest neighbors (NNN) leads to a more negative $\delta$($^{23}$Na) value. This result agrees well with the more negative shift of the $^{23}$Na resonance corresponding to Na in prismatic sites in P2$_1$/n NZC ($-11$~ppm) compared to P2$_1$/n NYC ($-8.1$~ppm). For NYZC75, slow cooling thus appears to yield a structure containing more Zr-rich prismatic Na environments, which is consistent with the fact that slow cooling tends to yield a more thermodynamically-equilibrated structure that more closely resembles the lowest energy NYZC75 supercell with Y and Zr clustering.   

To assess the impact of synthesis on Na-ion conduction properties, EIS was conducted on the three samples. Room temperature ionic conductivities for the twice ball milled and quenched samples (reported elsewhere\cite{wuStableCathodesolidElectrolyte2021}), as well as for the slow cooled sample, are $6.6\times10^{-5}$, $1.3\times10^{-5}$, and $9.5\times10^{-6}$~S~cm$^{-1}$, respectively, showing once again that an additional ball milling step greatly improves ionic conductivity. Specifically, the second ball milling step brings the activation energy for Na-ion transport down from 757~meV to 664~meV. To further assess the impact of hot pressing on the microstructure and conduction properties of twice ball milled NYZC75, cross-sectional SEM images and EIS data were also collected on a pellet densified under a pressure of 370~MPa at room temperature, and on a pellet that was hot pressed at 300~$^{\circ}$C under a similar pressure for 4~h. The hot pressed pellet, despite having a smaller void fraction than the room temperature densified pellet (see Figure~S20), exhibits a significantly lower conductivity of $2.3\times10^{-6}$~S~cm$^{-1}$ compared to $6.1\times10^{-5}$~S~cm$^{-1}$, consistent with the expected increase in crystallinity and reduced disorder after the heat treatment.



\begin{figure*}[htbp]
\centering
  \includegraphics[height=8.5cm]{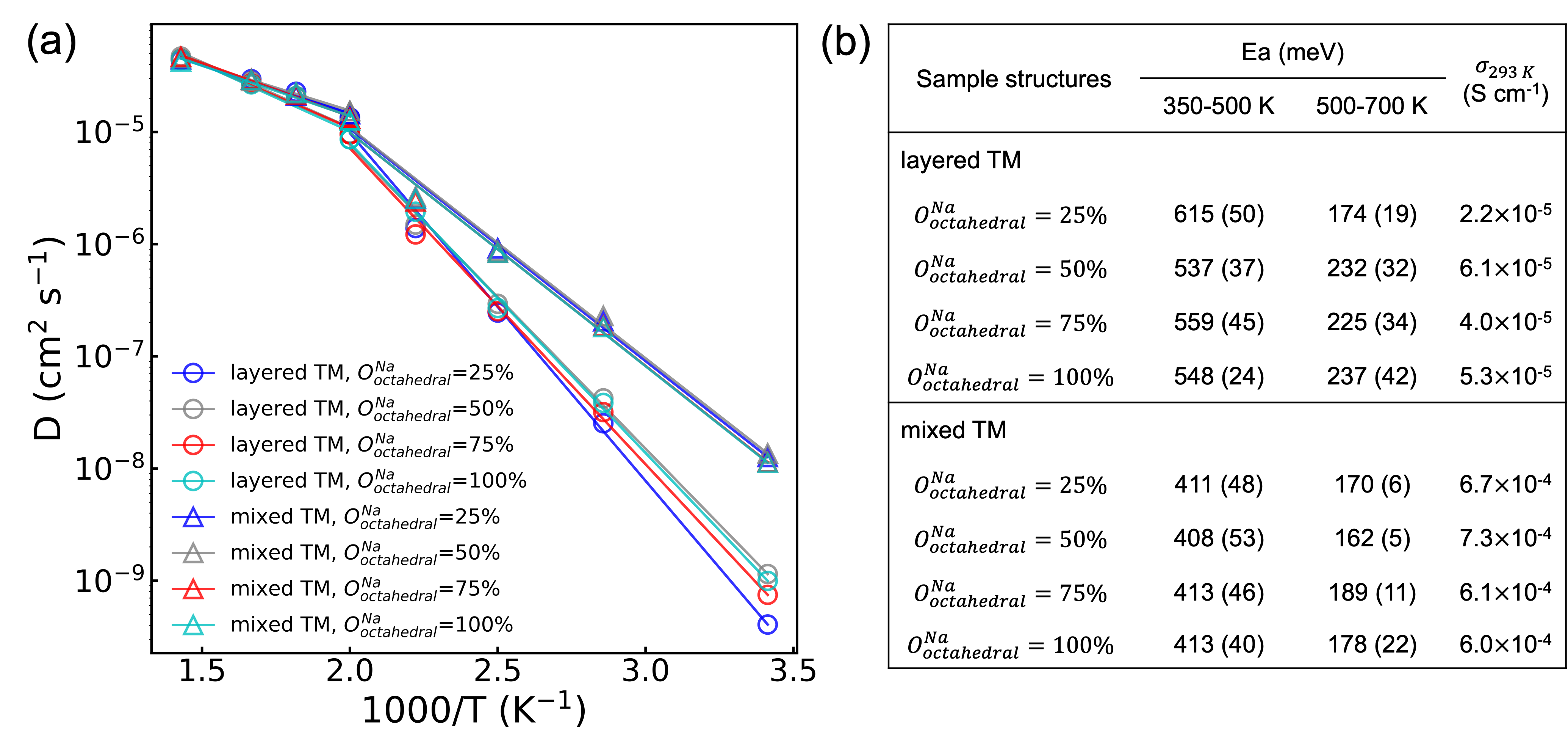}
  \caption{(a) Arrhenius plot of Na diffusivities obtained from MD simulations between 350 and 700~K for eight representative \ce{Na_{2.25}Y_{0.25}Zr_{0.75}Cl6} supercells spanning different Na octahedral occupancy values ($O^{Na}_{octahedral}$) and Y-Zr orderings (layered-TM and mixed-TM structures). The activation energies below and above the $\sim$ 500~K transition temperature, and the extrapolated $\sigma_{293~K}$ are listed in (b). Standard errors in the activation energies are provided in parentheses.}
  \label{fgr:Arrhenius and Ea table}
\end{figure*}

To probe the role of Na-vac and Y-Zr disorder on the conduction properties of NYZC75, molecular dynamics (MD) simulations were conducted between 350 and 700~K on eight NYZC75 structures spanning the two types of Y-Zr ordering, as well as four different $O^{Na}_{octahedral}$ values. To enable MD simulations of Na-ion transport on large NYZC75 unit cells and over long timescales, a moment tensor potential (MTP) was developed using an active learning training method and tested for accuracy and reliability on the structures of interest (see Figure~S21 and Methods section)\cite{shapeevMomentTensorPotentials2016, shapeevAccurateRepresentationFormation2017,novikovMLIPPackageMoment2021}. As shown in Figure \ref{fgr:Arrhenius and Ea table}, a transition between two quasi-linear Arrhenius regimes is observed for all of the structures at around 500 K. The low-temperature (350-500 K) activation energies ($E_a$) for the layered-TM structures range from 615 to 537 meV, in line with our previously-published computational results on the ground state NYZC75 structure\cite{wuStableCathodesolidElectrolyte2021}. The computed Na-ion conductivities at 293~K ($\sigma_{293~K}$) for the layered-TM structures range from 2.2 to 6.1$\times$10$^{-5}$~S~cm$^{-1} $, which is also consistent with previous simulation and experimental results\cite{wuStableCathodesolidElectrolyte2021}. 

Interestingly, the MD simulations predict significantly improved Na diffusion for the mixed-TM structures, with an average $E_a$ around 410 meV and $\sigma_{293~K}$ over 6.0$\times$10$^{-4}$~S~cm$^{-1}$. While a direct comparison between theoretically-predicted and experimentally-observed diffusion properties is complicated by the fact that the distribution of Y and Zr species on the transition metal lattice is difficult to ascertain experimentally, it is reasonable to expect that the twice ball-milled NYZC75 sample will lead to greater Y-Zr mixing than the quenched sample. Indirect evidence for greater mixing in the twice ball milled sample is provided by our $^{23}$Na ss-NMR results indicating the presence of a wider range of Na local environments in this sample. Given this, the trend observed by EIS is qualitatively in line with the MD predictions, whereby the activation energy measured on the twice ball milled sample (664 and 636~meV) is lower than that measured on the quenched sample (757~meV), and its $\sigma_{293~K}$ is greater. 

The MD simulation results presented in Figure~\ref{fgr:Arrhenius and Ea table} also indicate that the distribution of Na amongst available sites in the NYZC75 structure has a minimal effect on Na-ion diffusion. Samples with different $O^{Na}_{octahedral}$ values but identical Y-Zr SRO are found to have comparable E$_a$ and $\sigma_{293~K}$. By monitoring the evolution of Na site occupancies of the four layered-TM structures at 400~K (see Figure~S22a), we find that all of their $O^{Na}_{octahedral}$ converge to an average value of 32.5\%$\pm$1\% within 50~ps. This finding by MTP MD is consistent with our previous AIMD results that Na-ion transport occurs through hops between neighboring prismatic and octahedral sites (see Figure~S23). It is immediately clear that the initial Na site occupancies hardly affect the Na diffusion as they equilibrate once Na atoms start to diffuse. The equilibrated Na site occupancies, however, depend on to the Y-Zr SRO. As shown in Figure~S22b, average equilibrated $O^{Na}_{octahedral}$ values for mixed-TM structures ($\sim$ 37\%) is higher than that of layered-TM structures ($\sim$ 32\%) during the 10 ns MD production run at 400 K, indicating that TM mixing promotes Na site disorder as well. 

In addition to having consistently higher $O^{Na}_{octahedral}$, mixed-TM structures are also found to have partially activated Cl motion below the E$_a$ transition temperature at 500~K. As seen in Figure~S24, the isosurface of the Cl probability density at 400~K is broadened in the mixed-TM structures, indicative of partial Cl rotation. Conversely, the reduced Cl isosurface in the layered-TM structures suggests hindered Cl rotations. In our previous work\cite{wuStableCathodesolidElectrolyte2021}, we showed using AIMD simulations that Cl motion plays a significant role in promoting Na diffusion and decreasing $E_a$. Accordingly, the lower $E_a$ for Na-ion diffusion for mixed-TM structures likely results from partially activated Cl motion below 500~K enabled by greater Y-Zr mixing.   

\begin{figure*}
 \centering
 \includegraphics[height=9cm]{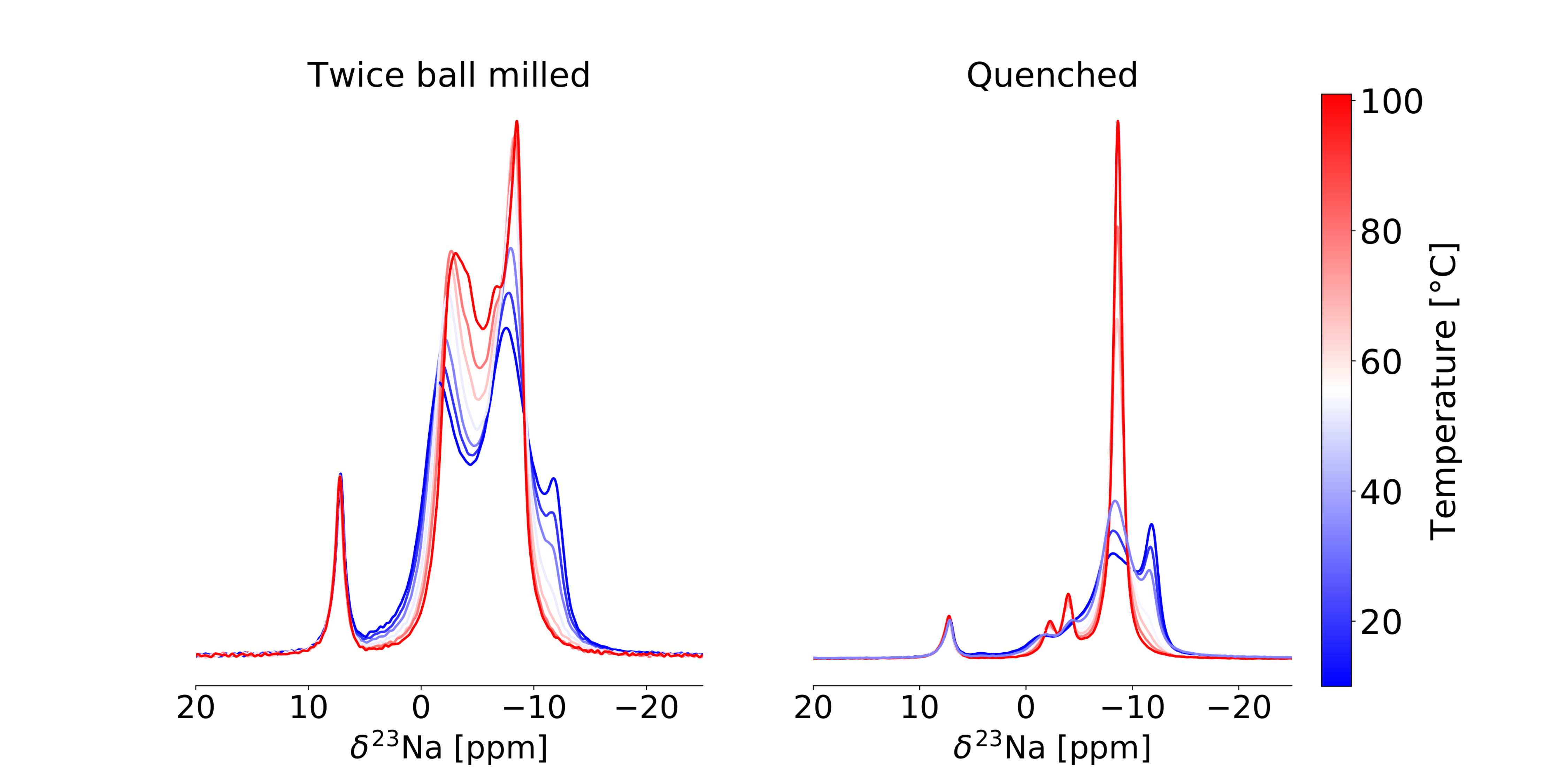}
 \caption{$^{23}$Na variable temperature ss-NMR spectra collected on the twice ball milled (left) and quenched (right) \ce{Na_{2.25}Y_{0.25}Zr_{0.75}Cl6} samples. Spectra are color-coded using the temperature scale to the right of the plots. The resonance at 7.2~ppm corresponds to NaCl. All spectra were obtained at 18.8~T and at a 10 kHz MAS speed.}
 \label{fgr:NYZC_VT_NMR}
\end{figure*}

$^{23}$Na VT-NMR was conducted on the quenched and twice ball milled samples to better understand the impact of crystallinity and Y-Zr disorder on Na-ion transport. 
In a first instance, the stability of the twice ball milled NYZC75 compound was assessed over the temperature range used in the VT-NMR measurements. For this, we carried out a two run EIS measurement (see Figure~S25), where the conductivity was probed every 20~$^{\circ}$C as the temperature was ramped up to 100~$^{\circ}$C, down to 20~$^{\circ}$C, and then back up to 100~$^{\circ}$C. Between the first and second run, the conductivity decreased slightly from 5.6 to $3.9\times10^{-5}$~S~cm$^{-1}$ and the activation energy increased from 637 to 685~meV, indicating that while some material evolution in this temperature range occurs, it only has a small effect on ion conduction. 
Changes in conductivity could be tied to an increase in crystallinity, particle size effects, or small changes in the Y-Zr distribution as revealed by a comparison of the $^{23}$Na NMR spectra taken on a twice ball milled sample before and after an overnight exposure to temperatures up to 111~$^{\circ}$C in the spectrometer (see Figure~S26). To reduce impact of these small structural changes on the $^{23}$Na VT-NMR results, we limited our temperature range to 10 to 101~$^{\circ}$C.
A superposition of the VT-NMR spectra collected on the twice ball milled and quenched samples is shown in Figure~\ref{fgr:NYZC_VT_NMR}. In the $^{23}$Na spectrum collected on the twice ball milled sample at 10~$^{\circ}$C, the distribution of Na environments differs slightly relative to the spectrum shown previously (Figure~\ref{fgr:NYZC75_NMR&CASTEP}a) and obtained at 52~$^{\circ}$C, with more intensity in the most negative resonance around $-13$~ppm. Upon increasing the temperature, the resonances in the range of $-13$ to $-10$~ppm rapidly lose intensity, indicating increased Na exchange involving those prismatic sites. As the temperature is further raised, all of the peaks move towards one another due to rapid Na exchange between all of the Na sites on the NMR timescale, corroborating a Prism-Oct-Prism hopping mechanism for Na-ion transport through the P2$_1$/n structure. Between 66 and 101~$^{\circ}$C, a sharp peak progressively grows at $-8.0$~ppm. 

For the $^{23}$Na spectrum collected on the quenched sample at 10~$^{\circ}$C, two resonances are observed at about $-8.0$ and $-13$~ppm, similarly to twice ball milled NYZC75 but with a more intense $-13$~ppm resonance consistent with a greater proportion of Na in prismatic, Zr-rich local environments in this sample. The prismatic resonance quickly loses intensity, feeding the rapidly growing and narrowing peak around $-8.0$~ppm that resembles the sharp peak observed at high temperature in the twice ball milled sample. Two smaller peaks grow and sharpen around $-2.0$ and $-4.0$~ppm, suggesting that these environments are also involved in the Na-ion conduction mechanism. Given that there are twice as many prismatic sites as octahedral sites, mobile Na-ions are expected to experience an average local environment that more closely resembles prismatic sites than octahedral sites as the temperature is increased. The sharp peak at $-8.0$~ppm is therefore attributed to rapidly diffusing Na occupying mostly prismatic environments. The narrow lineshape at $-8.0$~ppm indicates fast Na-ion motion through the entire structure, which agrees well with the high thermal factors for Na residing in prismatic sites in \ce{Na_{2.25}Er_{0.25}Zr_{0.75}Cl6}\cite{schlemNa3XEr12020}.


While our experimental and simulation results both indicate enhancements in Na diffusion with greater Y-Zr mixing, experimental conductivities are consistently lower than those predicted for the mixed-TM structures, and the activation barriers are consistently higher. These differences can be attributed to the likely limited Y-Zr disorder achievable \textit{via} ball milling. Meanwhile, our simulations and NMR characterization results provide information on the nature of Na and Y-Zr orderings present in each sample, but effects from structural amorphization and particle downsizing due to ball milling are not probed and could be significant. Finally, the overestimation of the lattice constants by the PBE functional\cite{qiBridgingGapSimulated2021} can be a potential further source of error. 
Nevertheless, the results presented here suggest that more promising Na electrolyte material may be developed through better synthetic control of transition metal distributions, potentially through the inclusion of a greater number of species as in high entropy systems, or through the selection of species that favor transition metal mixing over clustering. Moving forward, studies seeking to isolate and compare the effects of chemical distributions and structural amorphization on Na-ion transport are of particular interest.

\section{Conclusions}
In this work, we systematically explore the relationship between synthesis, structure, and Na-ion transport in the \ce{Na_{3-x}Y_{1-x}Zr_{x}Cl6} family of solid electrolytes, focusing on the influence of polymorphism in the end-member compounds and disorder in the \ce{Na_{2.25}Y_{0.25}Zr_{0.75}Cl6} composition. In the \ce{Na3YCl6} and \ce{Na2ZrCl6} systems, we leverage ss-NMR and DFT calculations to show that both compounds are subject to polymorphism, with phase compositions that can be tuned by varying synthesis/materials processing conditions. 
While previous studies have attributed the increased ionic conductivity of ball milled \ce{Na2ZrCl6} to disorder in the P$\bar{3}$m1 structure, we demonstrate the existence of a more conductive P2$_1$/n \ce{Na2ZrCl6} polymorph with isotropic Na-ion migration pathways and intrinsic vacancies.
For \ce{Na3YCl6}, non-equilibrium synthesis methods like quenching and ball milling result in the formation of the P2$_1$/n polymorph while slow cooling leads to the thermodynamically stable R$\bar{3}$ phase. Ball milling is necessary for improving the conductivity of the R$\bar{3}$ polymorph, likely through a reduction in crystallinity and the introduction of Y site disorder, while Na-ion transport within the P2$_1$/n phase is expected to be hindered by the lack of Na intrinsic vacancies.
In \ce{Na_{2.25}Y_{0.25}Zr_{0.75}Cl6}, ball milling leads to greater Y-Zr mixing, which greatly enhances its ionic conductivity. Computational simulations suggest that the conductivity improvements from better Y-Zr mixing are linked to activated Cl rotations about the \ce{YCl6^{3-}} and \ce{ZrCl6^{2-}} octahedra. 
Lessons learned from this work are expected to be generalizable to other halide solid electrolytes. Specifically, the energetic similarity of halide structures with vastly differing Na-ion transport properties requires careful control of the phases formed and of the extent of disorder introduced on the cation lattice in pursuit of improved catholyte compositions for high voltage SSSBs. The optimization of transition metal orderings in aliovalently doped ion conductors is a new direction of study for solid-state ion conductors that has yet to be explored in great detail.

\section{Experimental methods}

\subsection{Synthesis} 
All solid electrolyte materials were prepared in an Ar-filled glovebox (\ce{H2O} and \ce{O2} < 0.5 ppm), unless indicated otherwise. Stoichiometric mixtures of \ce{NaCl} (Sigma Aldrich), \ce{YCl3} (Sigma Aldrich), and \ce{ZrCl4} (Sigma Aldrich) were hand mixed using a mortar and pestle before being loaded into 50 mL zirconia ball milling jars along with eleven 10~mm diameter zirconia grinding media. Samples were ball milled (Retsch Emax) for 2~h at 500 rpm. Following milling, the powders were extracted, loaded into quartz ampoules, and flame sealed under vacuum. Next, the quartz tubes were annealed at 500~$^{\circ}$C for 24~h, after which they were quenched in a water bath. High-crystalline phases were obtained by slow cooling the powder product over 48 h, rather than water bath quenching. Lastly, to produce samples with low crystallinity, the annealed and quenched powders were re-loaded into  ball mill jars along with eighty-eight 5~mm diameter grinding media and milled at 400 rpm for 4~h.

\subsection{Materials characterization}
Powders for X-ray diffraction measurements were flame-sealed in 0.5 mm diameter boron-rich capillary tubes (Charles Supper). A Bruker X8-ApexII CCD Sealed Tube diffractometer was used to collect the X-ray diffraction patterns using a molybdenum source radiation ($\lambda_{Mo}$ = 0.7107~\AA) over a 5 – 50~$^{\circ}$ 2$\theta$ range with a step size of 0.01°. A FEI Scios DualBeam focused-ion beam scanning electron microscope (FIB-SEM) was used to collect SEM images of the solid electrolyte powders. To prevent air exposure of the solid electrolyte samples, an air-tight holder was used to load the powders within the glovebox, and then the samples were transferred to the SEM. Raman spectra were collected using a Renishaw inVia upright microscope and a 532~nm laser source. 

\subsection{Nuclear magnetic resonance}

All $^{23}$Na spectra were acquired at 18.8 T (800 MHz for $^1$H) on a Bruker Ultrashield Plus standard bore magnet equipped with an Avance III console using 2.5 mm and 3.2 mm HX MAS probes. All spectra are referenced to a 1 M NaCl aqueous solution at 0.04~ppm\cite{makulskiMultinuclearMagneticResonance2019}. For the 1~M NaCl aqueous solution, 90~$^{\circ}$ flip angle pulses were 2.46 and 5.7~$\mu$s at 100~W for the 2.5~mm HX and 3.2~mm HX probes, respectively. Samples were packed and sealed into 2.5~mm and 3.2~mm zirconia rotors closed with a single Vespel cap and a polytetrafluoroethylene (PTFE) spacer to maintain an inert atmosphere. All sample preparation was conducted in an Ar-filled glovebox. A 2000~L~h$^{-1}$ flow of \ce{N2} gas on the rotor also helped to prevent air exposure.  

A direct excitation, single pulse sequence was used for each 1D measurement with a magic angle spinning speed of 10 kHz. Because $^{23}$Na is quadrupolar ($I = 3/2$), short 30~$^{\circ}$ flip angle pulses were used to obtain quantitative spectra (2.5~mm HX: 0.29~$\mu$s at 200~W; 3.2~mm HX: 0.95~$\mu$s at 100~W). Recycle delays of 40~s were applied during signal averaging to ensure complete relaxation of all spins. 

Spectra acquired at variable temperatures (VT) were conducted using the aforementioned acquisition parameters from 10-135~$^{\circ}$C. Sample temperatures were controlled with a combination of heating coils and a SP Scientific’s XR AirJet Sample Cooler. Calibrated rotor temperatures are based on measurements of the longitudinal relaxation time (T$_1$) of $^{79}$Br in solid KBr under identical conditions\cite{thurberMeasurementSampleTemperatures2009}. 

2D $^{23}$Na EXchange SpectroscopY (EXSY) measurements were conducted using the 3.2 mm HX probe at a 10~kHz spinning speed and a calibrated sample temperature of 328~K. For the end-member compositions, 2.8~$\mu$s pulses at 100~W (90$^\circ$ flip angle) were used with 1~s mixing times and 5~s recycle delays. For NYC EXSY measurements, 16 scans were acquired for each of the 512 slices. For NZC EXSY experiments, 8 scans were acquired for each of the 1024 slices. For the mixed Y-Zr composition, pulses were 1.85~$\mu$s at 300~W (90$^\circ$ flip angle) with a 50~ms mixing time and a 5~s recycle delay. In this measurement, 128 scans were acquired for each of the 64 slices.

\subsection{Electrochemical impedance spectroscopy}
Ionic conductivities were extracted from EIS measurements and their corresponding Nyquist plots. Pellet cells consisting of 10 mm diameter Ti | SE | Ti were assembled inside a polyether ether ketone (PEEK) die, where pellets were formed by densifying ~70 mg of SE powder at 370 MPa using a hydraulic press (Carver). The EIS spectra were collected using a Solartron 1260A impedance analyzer with a sinusoidal amplitude of 30 mV and a frequency range of 1 MHz to 1 Hz. Activation energies for various solid electrolytes samples were determined by linear regression of the $ln(\sigma~T)$ vs. $1000/T$ data, following the Arrhenius equation.

\subsection{Computational methods}
$^{23}$Na NMR parameters were calculated with CASTEP\cite{clarkFirstPrinciplesMethods2005}. The exchange-correlation term was approximated with the generalized gradient approximation by Perdew, Burke, and Ernzerhof\cite{perdewGeneralizedGradientApproximation1996} and relativistic effects were accounted for using the scalar-relativistic zeroth-order regular approximation (ZORA)\cite{yatesRelativisticNuclearMagnetic2003}. The as-supplied "on-the-fly" ultrasoft pseudopotentials were used in all calculations\cite{vanderbiltSoftSelfconsistentPseudopotentials1990}. NMR parameter calculations used the projector augmented-wave method (GIPAW)\cite{yatesCalculationNMRChemical2007, pickardAllelectronMagneticResponse2001}. Each calculation followed an identical procedure of single point energy convergence, geometry optimization, and isotropic shift convergence as reported here\cite{sebtiStackingFaultsAssist2022}. 

Convergence parameters for each set of calculations are as follows. For single point energy calculations, the convergence tolerance was set to 0.5~meV~atom$^{-1}$. Geometry optimization calculations used a 0.02~meV~atom$^{-1}$ energy convergence tolerance, a 0.05~eV~\AA$^{-1}$ maximum ionic force tolerance, a 0.001~\AA~maximum ionic displacement tolerance, and a 0.1~GPA maximum stress component tolerance. The convergence criterion for calculations of $^{23}$Na NMR isotropic shifts was set to 0.5~ppm. Converged planewave energy cut-off values and \textit{k}-point grids are reported in Table~S1.

The calculated isotropic $^{23}$Na shifts were converted into experimentally relevant values through the construction of a semi-empirical calibration curve. Isotropic shifts for a series of binary Na-containing compositions were calculated and correlated to an experimentally-measured shift, as previously reported\cite{sadocNMRParametersAlkali2011}. The final calibration curve and a table of experimental and calculated shifts can be found in Supplementary Information Section~S26.

The construction of NYZC75 structures, structural relaxation with density functional theory (DFT) and ab initio molecular dynamics (AIMD) simulations were all performed and discussed in our previous work\cite{wuStableCathodesolidElectrolyte2021}. In this work, we only performed DFT static calculations on the newly selected training structures, where Vienna ab initio simulation package (VASP) was used with the Perdew-Burke-Ernzerhof (PBE) generalized gradient approximation (GGA) as the exchange-correlation functional, in line with our previous work\cite{wuStableCathodesolidElectrolyte2021}. The methodology of the augmentation of training set with active learning are discussed in details below. 

In our previous works, the MTP formalism has been proved to accurately simulate the room temperature ionic conductivities of superionic conductors, including \ce{Li3YCl6}, \ce{Li7P3S11}, \ce{Li_{0.33}La_{0.56}TiO3} and NYZC75\cite{qiBridgingGapSimulated2021,wuStableCathodesolidElectrolyte2021}. Though passive learning MTPs trained with manually constructed training set are capable to study bulk crystals, the construction of a comprehensive training set becomes challenging for more sophisticated tasks, e.g., grain boundary structures, amorphous structures and co-existence of multiple types of chemical orderings in this work. To automate and optimize the selection of training structures, Podryabinkin and Shapeev propose an active learning scheme with an extrapolation grade $\gamma$ to evaluate the extent to which a given configuration is extrapolative with respect to the existing training structures, thereby correlating the prediction error without ab initio information\cite{shapeevMomentTensorPotentials2016, shapeevAccurateRepresentationFormation2017, novikovMLIPPackageMoment2021}. With this scheme, distinctive and representative structures, i.e., structures with large $\gamma$ emerged in target MD simulations are selected automatically, thus resulting in a robust training set for target simulation conditions. By contrast, the traditional scheme of manual selection of training structures involves more trails and errors while failing to avoid repetitive configurations, so that passive learning MTPs tends to be less reliable than active learning MTPs. 

In this work, we used active learning to improve our previous passive learning MTP trained specifically for the most stable NYZC75 structure\cite{wuStableCathodesolidElectrolyte2021}. 34 iterations of active learning were conducted from 300 to 1200 K with 300 K intervals for the 8 NYZC75 structures spanning 2 types of Y-Zr SRO and 4 values of Na site occupancy, until the MTP can reliably complete 100 ps of MD simulations, i.e., $\gamma$ of MD snapshots are all below five in the $34^{th}$ iteration. (see Figure~S21) Beside MD reliability, the root mean square error (RMSE) in reproducing the DFT energies of the 229 enumerated NYZC75 structures by the active learning MTP (2.32 meV~atom$^{-1}$) is also smaller than that of the passive learning MTP (3.01 meV~atom$^{-1}$) (see Figure~S18). We thereby use the active learning MTP to study the effect of Na site occupancies and Y-Zr SRO on the ionic conductivity of NYZC75. MD NPT simulations were performed at 350 to 700 K for the 2$\times$2$\times$2 supercells of the eight sample structures used in active learning. At each temperature, 5 parallel runs were performed for at least 2 ns, and the stepwise averaged mean square displacement (MSD) were used to calculate diffusivity. The respective analysis on MD trajectories to extract diffusivities and ionic conductivities were performed with the pymatgen-analysis-diffusion package. All training, active learning, evaluations and simulations with MTP were performed using MLIP, LAMMPS and maml.

The definition of the pairwise multicomponent SRO parameter is
\begin{equation}
    SRO_{ij} = \frac{p_{ij} - c_j}{\delta_{ij} - c_j}
\end{equation}
where $p_{ij}$ is the average probability of finding a j-type atom around an i-type atom, $c_j$ is the average concentration of j-type atom in the system, and $\delta_{ij}$ is the Kronecker delta function. For pairs of the same species (i.e., i = j), a positive $SRO_{ij}$ means self-clustering, and a negative value suggests self-segregation. For pairs of different elements (i.e., i $\neq$ j), it is the opposite. A negative $SRO_{ij}$ suggests the tendency of j-type atoms clustering around i-type atoms, while a positive value means the segregation of i- and j-type atoms. In this study, the cut-off radius of 8.5 \text{\AA} was used to define the neighborhood of an atom, which is in between the shortest and the second shortest bond length of Y-Y, Y-Zr and Zr-Zr pairs. Also, to effectively distinguish structures with same Y-Zr orderings, only Y and Zr atoms are considered in the analysis, while Na and Cl are ignored.

\section*{Author Contributions}
E.S., P.M.R., and R.G. contributed to NMR measurements; J.Q., E.S., and S.B. contributed to DFT and MTP simulations and analysis; P.R. and E.A.W conducted synthesis, XRD, and EIS characterization; A.C. and S-Y.H contributed the Raman and SEM measurements; R.J.C, S.P.O, and S.Y.M supervised the study; E.S., J.Q., and R.J.C wrote the manuscript and all authors revised the manuscript.


\section*{Conflicts of interest}
There are no conflicts to declare.

\section*{Acknowledgements}
This work made use of the shared facilities of the UC Santa Barbara MRSEC (Grant DMR 1720256), a member of the Materials Research Facilities Network (http://www.mrfn.org), and the computational facilities administered by the Center for Scientific Computing at the CNSI and MRL (an NSF MRSEC; Grants CNS 1725797 and DMR 1720256). Use was made of computational facilities purchased with funds from the National Science Foundation (Grant CNS-1725797) and administered by the Center for Scientific Computing (CSC). The CSC is supported by the California NanoSystems Institute and the Materials Research Science and Engineering Center (MRSEC; NSF Grant DMR 1720256) at UC Santa Barbara. This material is based upon work supported by the National Science Foundation Graduate Research Fellowship under Grant NSF DGE 1650114. S.P.O. and S.B., and J.Q. would like to acknowledge funding provided by NSF FMRG under award number 2134764. The DFT and MD studies was performed using the Extreme Science and Engineering Discovery Environment (XSEDE, which is supported by National Science Foundation Grant ACI-1053575).



\balance


\bibliography{References} 
\bibliographystyle{rsc} 

\end{document}